\pdfoutput=1
\documentclass[iop]{emulateapj}
\usepackage{epsfig}
\usepackage{csquotes}
\usepackage{graphicx}
\usepackage{ulem,color}
\usepackage[dvipsnames]{xcolor}
\usepackage{dcolumn}
\usepackage{bm}

\usepackage{natbib}
\usepackage{hyperref}
\usepackage{xspace}
\usepackage{amsmath}

\usepackage{float}

\def\lesssim{\mathrel{\hbox{\rlap{\hbox{\lower4pt\hbox{$\sim$}}}\hbox{$<$}}}}
\def\gtrsim{\mathrel{\hbox{\rlap{\hbox{\lower4pt\hbox{$\sim$}}}\hbox{$>$}}}}

\newcommand{\bea}{\begin{eqnarray}}
\newcommand{\eea}{\end{eqnarray}}

\newcommand{\beq}[1]{\begin{equation} #1 \end{equation}}

\newcommand{\deriv}[2]{\frac{ d #1 }{ d #2 }}
\newcommand{\pderiv}[2]{\frac{ \partial #1 }{ \partial #2 }}
\newcommand{\harm}{{\sc Harm3d}\xspace} 
\newcommand{\bothros}{{\sc Bothros}\xspace} 
\newcommand{\Rmax}{r_\mathrm{max}}

\newcommand{\msun}{M_\odot}

\newcommand{\Omegabin}{\Omega_\mathrm{bin}}

\newcommand{\kev}{\mathrm{keV}}
\newcommand{\rmax}{\Rmax}
\newcommand{\rcam}{r_\mathrm{cam}}
\newcommand{\thcam}{\theta_\mathrm{cam}}
\newcommand{\phcam}{\phi_\mathrm{cam}}

\newcommand{\Teff}{T_\mathrm{eff}}

\newcommand{\rhori}{r_{\mathrm{hor} \, i}}

\newcommand{\lt}{\ensuremath <}

\def\lambdabar{%
\relax
\bgroup
\def\@tempa{\hbox{\raise.73\ht0
\hbox to0pt{\kern.25\wd0\vrule width.5\wd0
height.1pt depth.1pt\hss}\box0}}%
\mathchoice{\setbox0\hbox{$\displaystyle\lambda$}\@tempa}%
{\setbox0\hbox{$\textstyle\lambda$}\@tempa}%
{\setbox0\hbox{$\scriptstyle\lambda$}\@tempa}%
{\setbox0\hbox{$\scriptscriptstyle\lambda$}\@tempa}%
\egroup
}

\newcommand{\mdotlowvalue}{8 \times 10^{-4}}
  
\newcommand{\mdothighvalue}{0.5} 

\newcommand{\mdotlow}{\dot{m}=\mdotlowvalue} 
 
\newcommand{\mdothigh}{\dot{m}=\mdothighvalue}

\slugcomment{Submitted to ApJ.}

\begin{document}

\title{Electromagnetic Emission from Supermassive Binary Black Holes Approaching Merger}

\author{St\'ephane d'Ascoli    $^{1,2}$,
		Scott C. Noble       $^{3,4}$,
		Dennis B. Bowen      $^1$,
        Manuela Campanelli   $^1$,
        Julian H. Krolik     $^5$,
        and Vassilios Mewes  $^1$}

\affil{$^1$ Center for Computational Relativity and Gravitation, 
  Rochester Institute of Technology, Rochester, NY 14623, USA\\
  $^2$ \'Ecole Normale Sup\'erieure, 24 rue Lhomond, 75005 Paris, France\\
  $^3$ Department of Physics and Engineering Physics,
  The University of Tulsa, Tulsa, OK 74104, USA\\
  $^4$ Gravitational Astrophysics Laboratory, NASA Goddard Space Flight Center, Greenbelt, MD 20771, USA\\
  $^5$ Department of Physics and Astronomy, Johns Hopkins
  University, Baltimore, MD 21218, USA\\
  }

\email{scott.c.noble@nasa.gov}

\begin{abstract}
We present the first fully relativistic prediction of the electromagnetic emission
from the surrounding gas of a supermassive binary black hole system approaching merger.
Using a ray-tracing code to post-process data from a general
relativistic 3-d MHD simulation, we generate images and spectra, and
analyze the viewing angle dependence of the light emitted.  When the
accretion rate is relatively high, the circumbinary disk, accretion
streams, and mini-disks combine to emit light in the UV/EUV bands.  We
posit a thermal Compton hard X-ray spectrum for coronal emission; at
high accretion rates, it is almost entirely produced in the
mini-disks, but at lower accretion rates it is the primary radiation
mechanism in the mini-disks and accretion streams as well. Due to
relativistic beaming and gravitational lensing, the angular
distribution of the power radiated is strongly anisotropic, especially
near the equatorial plane.
\end{abstract}


\section{Introduction} 

\subsection{Context}

Electromagnetic (EM) observations of supermassive binary black holes
(SMBBHs) and their environments have the potential to provide critical
new information about both galaxy evolution and strong-field
gravity. Unlike merging stellar-mass binary black holes (BBHs)
recently discovered by the LIGO-Virgo Collaboration (LVC)
~\citep{TheLIGOScientific:2016qqj,Abbott:2016nmj,TheLIGOScientific:2016pea,2017PhRvL.118v1101A},
SMBBHs may often merge in gas-rich environments \citep{Dotti09,Pfister17,Goulding18} and can
therefore be EM-bright during all stages of the coalescence process.

Direct detection of SMBBHs through gravitational wave (GW) emission
may be accomplished by orbiting GW observatories, but not any time
soon \citep{lisa-proposal2017}.  Pulsar
Timing Array (PTA) observations could detect GW radiation from SMBBHs,
but the GW frequencies to which they are sensitive correspond only
to the weakly relativistic regime for the most massive SMBBHs
$\left( \gtrsim 10^{9} \msun\right)$ \citep{SHANNON15}.  Identification
of {\it photons} from SMBBHs by some of the many EM telescopes operating now
could jump-start this field, sharply refining our estimates of the population and
evolution of SMBBHs, as well as guiding planning and development of
space-based GW observatories.

Observational efforts to identify SMBBHs to date have been defined by qualitative guesses
about what observable properties might be distinctive. One approach
to finding true SMBBHs
is focused on high-resolution imaging, possible only via radio-frequency
Very-Long-Baseline Interferometry (VLBI) \citep{Rodriguez09,Tremblay16}.
The recent discovery of possible orbital motion in radio galaxy 0402+379 presents an exciting new prospect of probing a
SMBBH's kinematics \citep{2017ApJ...843...14B}.
Another approach rests upon the hope that some
aspects of their light may exhibit periodic variability \citep{Graham2015,Liu2015,Liu2016}.
The latter approach is made difficult by the fact that the monitoring
programs rarely cover more than a few cycles of the candidate periods identified,
thus providing only weak evidence for periodicity.  
The results presented here are a first step toward
establishing more physically-grounded predictions of distinctive
spectral and timing properties of these intriguing systems.

\subsection{Prior Work}

The structure of a circumbinary disk when the binary mass ratio
$q \equiv M_2/M_1 \gtrsim 0.02$ is well-established: if the binary
semi-major axis is $a$ and its eccentricity is $e$, a gap forms
within a radius from the center-of-mass $\approx 2a(1+e)$ because
closed orbits enclosing the binary do not exist at smaller radii,
while an ordinary accretion disk occupies radii $\gg 2a$ \citep{P91,MM08,Roedig:2011,Shi12}.
Although early work treating a 1-d model of such a system argued that
torques exerted by the binary would prevent any accretion through
the gap \citep{P91}, detailed 2-d and 3-d
simulations have shown that,
although matter can pile up near $r \approx 2a$, ultimately inflow
equilibrium is achieved so that the mass accretion rate across the
gap matches that in the outer parts of the circumbinary disk
\citep{MM08,Shi12,Noble12,Roedig:2012,DOrazio13,Farris14,Shi15}.  Matter
crosses the gap in narrow streams, whose ultimate destinations
depend upon the matter's specific angular momentum.  Gas with specific
angular momentum close to the circular orbit value at the circumbinary
disk's inner edge spend enough time in the gap that binary torques
propel them back to $r \approx 2a$, where they shock against the
disk; gas with significantly lower angular momentum is created by
deflection in these shocks and plunges into the zone of the binary
\citep{Shi15}.  Once the streams find their way close to the binary, they
join one of the ``mini-disks", individual accretion disks each centered on
one of the partners in the binary \citep{RyanMacFadyen16,Bowen17,2017MNRAS.469.4258T,Bowen18}.
Although it was initially expected that accretion from the circumbinary
disk to the mini-disks would be cut off when the timescale on which the
binary orbit shrinks due to gravitational wave radiation becomes shorter
than the accretion timescale in the inner region of the circumbinary disk
\citep{MP05}, simulations have shown that this cut-off does not actually
occur \citep{Noble12,Farris15} because the very shortness of the binary
lifetime means that only material very close to the edge of the circumbinary
disk needs to be drawn upon to feed the flow.

Radiation can arise in any of these regions: the circumbinary disk, the
streams, and the mini-disks.  Its energy source is dissipation
of kinetic energy and magnetic field energy, which, in turn, is drawn
from mass moving into deeper portions of the gravitational potential.
Several different mechanisms can contribute to this dissipation.  In
ordinary accretion disks, the majority of the heat is generated by
dissipation at the short lengthscale end of the inertial cascade
associated with magnetohydrodynamic (MHD) turbulence stirred 
by the magnetoroational instability 
(MRI)~\citep{BH98}.  A smaller portion can be generated by magnetic
reconnection and related effects in the atmospheres of disks, their
``coron\ae" \citep{NK09}.  These two processes are responsible for most of the
dissipation in the circumbinary disk and possibly in the mini-disks.
In addition, however, in the context of binary accretion, shocks can
contribute in several ways.  There are the shocks already mentioned,
where outward-moving streams strike the inner edge of the circumbinary
disk; their luminosity has been previously discussed in
\cite{Noble12,Tang:2018rfm}.  There can also be shocks where inward-moving streams strike
the outer edge of a mini-disk \citep{Roedig:2014,Farris15b}.  If the mini-disks are sufficiently
hot (sound speed $c_s$ at least $\sim 0.1 v_{\rm orb}$, for $v_{\rm orb}$
the speed of a circular orbit at the relevant location), tidal interactions
and stream impacts can generate spiral shocks of substantial amplitude
within the mini-disks \citep{Ju16,RyanMacFadyen16,Bowen17}.

Radiation can also be created in the course of the merger proper, but most calculations
of it so far have been conducted only at the level of ``proof of principle"
\citep{2010ApJ...715.1117B,Pal10,Farris11,Bode12,
Farris12,Giacomazzo12,Farris14,Gold14, Kelly:2017xck}.  Because these were fully
relativistic calculations (i.e., they compute the changing spacetime
simultaneously with evolving the fluid),
the exceedingly high computational costs made it impossible to integrate
long enough for accretion to fill the mini-disks, even though Nature
would have had ample time to do so.

More has been accomplished concerning the epoch of approach to merger.
Using analytic estimates for disk structure, \citet{Roedig:2014} argued
that when the binary separation is at least several tens of gravitational
radii ($r_g \equiv GM/c^2 = M$ for $G=c=1$; here $M$ is the total
binary mass), but close enough that the
circumbinary disk is able to radiate a significant luminosity (i.e.,
$a \lesssim 300M$), there should be a ``notch" in the thermal
spectrum due to the weakness of radiation from the accretion streams crossing
the gap.  This notch might appear anywhere from the near-IR to the near-UV,
depending on parameters. They further predicted that in this phase of SMBBH evolution
there should be a substantial hard X-ray component due to Compton cooling
of the gas shocked when an accretion stream strikes the outer edge of
a mini-disk.

A number of papers adopting 2-d hydrodynamics and assuming both accretion
stress and dissipation are described by a phenomenological ``$\alpha$"
viscosity have explored accretion onto mini-disks over long enough periods
of time for the circumbinary disk to reach equilibrium out to large
multiples of $a$ \citep{MM08,Farris14,Farris15,RyanMacFadyen16,Tang:2018rfm}.
In all but one of the papers of this group, the fluid dynamics were Newtonian,
and took place in a Paczynski-Wiita potential; in \citep{RyanMacFadyen16},
a single mini-disk was studied with 2-d GR hydrodynamics
in a Schwarzschild spacetime perturbed by Newtonian tidal gravity to
approximate the influence of the other black hole (BH).

These simulations have yielded predictions of the emitted radiation,
doing so by describing the cooling rate in terms of their $\alpha$ parameter
combined with a disk dynamical temperature defined ignoring radiation
pressure and scaled by an assumed disk Mach number (again, \citet{RyanMacFadyen16} is the
exception: in this paper the surface brightness in the fluid rest-frame is
defined to be the thermal rate at a temperature defined by the ratio of vertically-integrated
gas pressure to surface density, but divided by the vertical optical depth).
A principal result of this series of papers is the prediction of a
spectrum comprising three quasi-Planckian thermal peaks, one at $\simeq 1$~keV arising
principally from the circumbinary disk, another (somewhat weaker) component
at $\simeq 3$~keV emitted by the streams, and a third at $\simeq 20$~keV
radiated by the mini-disks, but fading over time \citep{Tang:2018rfm}.

Recently, 
\citet{Bowen17, Bowen18} reported the first simulations
of mini-disk dynamics when the binary separation is small enough (a
few tens of $M$) that the orbit evolves due to GW emission.
Using a fully-relativistic spacetime for a binary comprising an equal-mass
pair of non-spinning BHs, they encountered several surprises.
Because the relativistic gravitational
potential between the two BHs becomes shallower than in the Newtonian
regime, the mini-disks stretch toward the L1 point and the amount of
gas passing---or ``sloshing''---back and forth between them
increases sharply when the separation is $\lesssim 30M$. The sloshing is
quasi-periodically modulated at a frequency $\simeq 2$--2.75 $\Omegabin$,
where $\Omegabin$ is the binary's orbital frequency. Although tidal
effects in Newtonian binaries are known to induce $m=2$ spiral waves
in mini-disks, the leading-order post-Newtonian (PN) 
terms induce strong $m=1$ features.
Perhaps most surprisingly, when the separation is as small as $20M$,
the inflow time in the mini-disks is so short that their mass
responds strongly to modulation of their supply rate on the binary orbital
timescale.

\subsection{Our Work}
Here we will make use of the data reported in \cite{Bowen18}, 
produced using the \harm code \citep{Noble09}, to make
detailed predictions of both the spectrum and the time-dependence of
the light emitted.  Because the \harm code  
is both intrinsically
conservative and uses a local cooling function to radiate nearly all
the heat produced, whether generated by turbulent dissipation or shocks,
the luminosity we predict is automatically consistent with the energy
budget of the flow.  These predictions are, however, dependent
upon two assumptions about the fluid-frame spectrum: where the gas is optically thick, we assume
it radiates a local black-body spectrum; and where the gas is optically thin,
we assume it radiates hard X-rays in a manner similar to AGN, emitting
a thermal Compton spectrum with temperature $kT = 0.2 m_e c^2 \simeq 100$~keV.

The remainder of this paper is organized as follows. In
Section~\ref{sec:methodology}, we specify the means by which we
determine the EM emission from the simulation data and transport it
through the binary's spacetime to a simulated observer.  Then in
Section~\ref{sec:results}, we describe the results of our ray-tracing
calculations.  In Section~\ref{sec:discussion} we discuss the
implications of our findings, and summarize them
in Section~\ref{sec:conclusion}.

\section{Methodology}
\label{sec:methodology}

Calculating the radiation observed at infinity produced by gas in
the state determined by the simulation
requires a number of steps, and the use of different codes
and techniques.  In Section~\ref{sec:grmhd-simul-deta}, we briefly
describe the numerical details and assumptions behind the simulation;
further details are given in a separate paper focused
on its analysis~\citep{Bowen18}.  We then explain our model for the
thermodynamics assumed in the simulation in
Section~\ref{sec:thermodynamic-model} and provide the
specifics of the radiative transfer solution in
Section~\ref{sec:ray-tracing-method}.  

Space and time coordinates are reported in units of
the total BBH mass, $M$, assuming geometrized units $G=c=1$.  Stated
times of snapshots are elapsed times from the start of
the simulation, generally quoted in terms of the initial
binary orbital period $t_{\rm bin} \simeq 600M$.

\subsection{GRMHD Simulation Details} 
\label{sec:grmhd-simul-deta}

Our calculation assumes the existence of an accretion flow around
an inspiraling equal-mass binary with an initial separation $a_0 = 20M$.
In the immediate vicinity of each BH, the spacetime is well
modeled as a boosted and perturbed BH
spacetime~\citep{Poisson:2005pi,Detweiler:2005kq} in
horizon-penetrating coordinates~\citep{JohnsonMcDaniel:2009dq}.
Elsewhere in the domain, the spacetime may be described with PN
theory~\citep{Blanchet:2014av}. The PN solution is combined with black
hole perturbation theory via asymptotic matching in regions of common
validity to produce a global analytic spacetime, which has been shown
to reasonably satisfy the Einstein Field Equations and is described in
full detail in~\citep{Proj0,ProjS0}.  The entire spacetime is
described in terms of PN Harmonic (PNH) coordinates, which are also
the coordinates used in our ray-tracing calculations.  The PN
description is sufficiently high-order to self-consistently contain
both gravitational radiation and the consequent orbital evolution.

The MHD simulation was performed using the \harm code~\citep{Noble09},
which evolves the magnetized matter on a background spacetime (which can
be, as it is here, time-dependent) through conservation of baryon number
density, conservation of stress-energy, and the Maxwell induction
equation (see~\citet{Noble09} for more details).  In the initial state,
a circumbinary disk (whose properties are taken from the quasi-steady
state $t=50,000M$ snapshot of the {\tt RunSS} simulation described in
\citet{Noble12}) occupies the region outside $\approx 2a$ from the
center-of-mass, while identical mini-disks fill the BHs'
Roche lobes, each threaded by a weak poloidal magnetic field (that is,
poloidal with respect to the central BH and a polar axis parallel to the
orbital axis).  If there were no magnetic field, both disks would be
close to dynamical equilibrium.  The magnetic field eventually destabilizes
the disks through magnetic winding and the development of the MRI;
the resulting magnetic stress helps the disks accrete and evolve.  Details
regarding the mini-disk construction can be found in~\cite{Bowen17,Bowen18}.

\subsection{Thermodynamic Model}
\label{sec:thermodynamic-model}

Because the simulation's flux-conservative numerical methods ensure
that all dissipated energy is captured and turned into heat, we are
able to self-consistently predict the amount of light emitted by our
simulation.  If there were no losses, the heat retained by the
fluid would continually add vertical pressure support in the disk and
geometrically thicken it.  The thermal energy would then be accreted
into the BH or be carried out from the disk by a wind.  By adding a
loss term to the energy equation, we can create a more realistic
structural model for the accretion flow (i.e. disks with constant
aspect ratio supported by gas pressure) while also evaluating the
amount of energy available for radiation.

This loss term mimics a bolometric (frequency-integrated) cooling rate
and appears as a sink term in the gas's equation of motion:
$\nabla_{\lambda}{T^{\lambda}}_{\mu} = -\mathcal{L}_c u_{\mu}$.
Its recorded value serves as a bolometric source to our radiation
transfer solution, which is performed as a post-processing step.
This procedure contrasts with other post-processing methods in
which the gas's emissivity is determined by the temperature found
in the simulation without consideration of radiative losses
(e.g., \cite{2010ApJ...715.1117B,Farris11,Kelly:2017xck}).

As in~\citet{Noble12}, the cooling function is designed to cool the gas
towards the initial entropy ($S_0 = 0.01$) at a rate per unit volume
\begin{equation}
  \mathcal{L}_{c} = \frac{\rho \epsilon}{t_{\rm
  cool}} \left( \frac{\Delta S}{S_0} + \left| \frac{\Delta
  S}{S_0}\right| \right)^{1/2} \ ,
\end{equation}
where $\rho$ is the rest-mass density, $\epsilon$ is the specific
internal energy, $\Delta S \equiv S - S_0$, and $t_{\rm cool}$ is the cooling
timescale. 
Choosing the target entropy as its initial value allows
us to measure how much total heat is produced by internal dissipation from the
simulation's onset to its completion.

We set four distinct regions of cooling timescales
following~\cite{Bowen17}.  In the outermost region, $r \ge 1.5a$, the gas
is cooled over a time equal to the the local
Keplerian orbital period: $t_{\rm cool} = 2\pi \left(r +
M\right)^{3/2}/\sqrt{M}$. Here, $r$ is the PNH radial
coordinate distance from the
center-of-mass~\citep{Blanchet:2014av}. In the vicinity of an
individual BH of mass $m_i$, where $r_{i} \leq 0.45a$, the cooling
time is set using the local Boyer-Lindquist coordinates:
$t_{\rm cool}= 2\pi r_{BL}^{3/2}/\sqrt{m_i}$.
For the coordinate mappings relating
the PNH coordinate system to the local Boyer-Lindquist coordinates, please
see~\cite{Bowen17}.  In the remaining portion of the simulation domain,
between the mini-disks and the circumbinary disk, the cooling time
is set to the value found at the inner edge of the outer region, 
$t_{\rm cool}(r=1.5a)$.

\begin{figure}[htb]
\center
\includegraphics[width=\columnwidth]{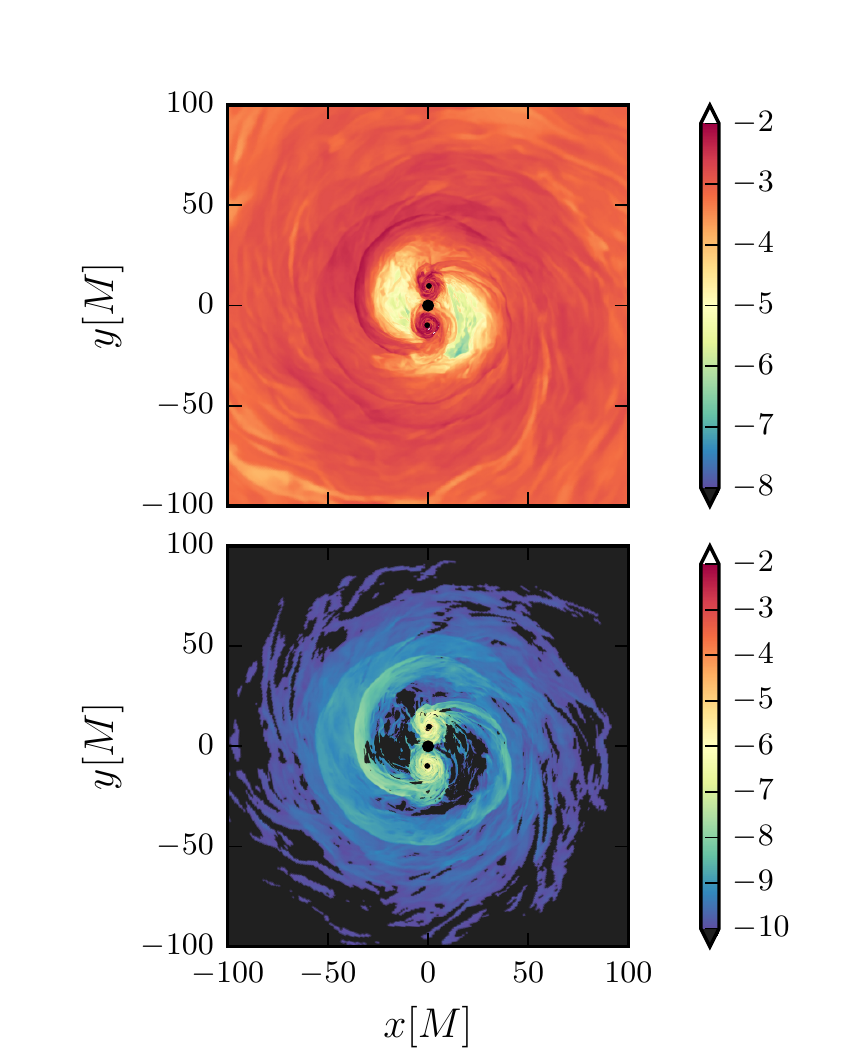}
\caption{Snapshot at $t=1030M$ of the rest-mass density $\rho$ (top) 
and cooling function $\mathcal{L}_c$ (bottom) in the equatorial plane
of the \harm simulation, using a logarithmic color scale for each.
The BHs' horizons are denoted as black circles displaced from the
origin, while the black circle in the center represents the coordinate
cutout at the origin. 
The horizontal and vertical coordinates are in the PNH Cartesian
coordinates.  We have set $\mathcal{L}_c = 0$ (black) where it is ignored
in the radiative transfer calculation. }
\label{fig:eq_snaps}
\end{figure}

\begin{figure}[htb]
\center
\includegraphics[width=\columnwidth]{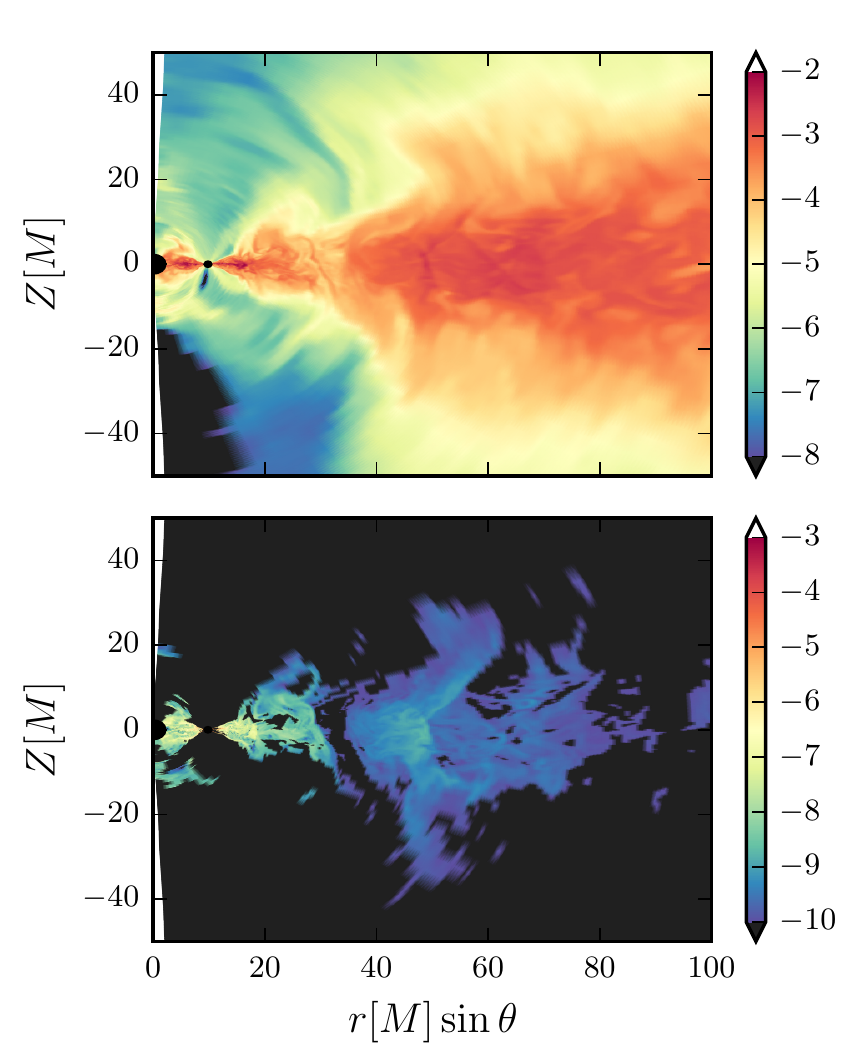}
\caption{Snapshot at $t=1030M$ of the rest-mass density $\rho$ (top) 
and cooling function $\mathcal{L}_c$ (bottom) in a poloidal plane
through one of the BHs; the color scale is logarithmic for
both quantities.  The BHs' horizons are denoted as black circles
displaced from the origin, while the black circle in the center
represents the coordinate cutout at the origin.  The horizontal and
vertical coordinates are in the PNH Cartesian coordinates.  
We have set
$\mathcal{L}_c = 0$ (black) where it is ignored in the radiative
transfer calculation.}
\label{fig:pol_snaps}
\end{figure}

Sample density and cooling function distributions in the
equatorial plane as well as in a poloidal slice through one of the
BHs are shown in Figures~\ref{fig:eq_snaps}~and~\ref{fig:pol_snaps},
respectively. To avoid contamination from excessively bright regions
caused by artificially high-entropy zones in floor states, our
ray-tracing calculations neglect emission from fluid cells with $\rho
< 10^{-4}$ in code units.  This density scale was consistently identified
throughout the simulation's duration to be  where gas
transitioned from a dense, turbulent state characteristic of a disk to a more
laminar, tenuous flow characteristic of a corona.
The ``cutout'' region covering the polar-coordinate origin at
the center-of-mass appears in the center in black in the 
plots of Figures~\ref{fig:eq_snaps}~and~\ref{fig:pol_snaps} (see
Section~\ref{sec:interpolation} for further details). 

\subsection{Ray-tracing Method}
\label{sec:ray-tracing-method}
\label{sec:bothros}

After the simulation data has been generated, it is post-processed
using a general relativistic ray-tracing code
called~\bothros~\citep{Noble07}.  The code has been used to calculate
the EM emission from a variety of single BH accretion
simulations~\citep{NK09,Noble09,Noble11} and is used here for the first
time in a dynamical spacetime.  We provide a brief summary of the code
before continuing with a description of the new aspects necessary for the
work presented here.

\bothros allows a user to produce time and frequency
dependent images of gas and is specifically tailored to systems
including single and binary BHs.  It approximates radiation as
freely-moving light rays---or null-like geodesics---within the system's
curved spacetime.  The code uses an observer-to-source approach,
shooting photons from a distant pinhole camera in various directions
through the source volume.  For a fixed camera location, tracing photons
backward is advantageous computationally, as only the light rays received 
by the observer are calculated.  Each ray
that is launched ultimately contributes a spectrum, $I_\nu$, to each pixel in
the simulated camera. The camera can be positioned at any point in
space, often specified in spherical coordinates
$\left\{\rcam, \thcam, \phcam \right\}$.  The integral of $I_\nu$ over
the pixels produces the locally-imaged flux spectrum $F_\nu(\rcam,\thcam,\phcam)$.

If the integrated optical depth along a ray never reaches unity, the
geodesic is terminated when it either exits the simulation domain
or reaches a distance $r_i < 1.001 \, \rhori$
from the $i^\mathrm{th}$ BH with horizon radius $\rhori$.
Otherwise, the geodesic is terminated at the photosphere.  From either
kind of termination point, the radiation transfer equation is
integrated along the geodesic in the opposite direction back to the
camera.
Its efficiency is large enough that, using 16
cores, \bothros is capable of producing images with resolution
matching the simulation resolution in only a few minutes per frequency
per simulation snapshot.  The ability to process 3-d time-dependent
GRMHD data efficiently gives the code an advantage over other codes
that rely on analytic models or angle-averaged data
(e.g., \cite{2012MNRAS.424.2504Z}) and more time-consuming (though
more versatile) Monte-Carlo approaches~\citep{2013ApJ...777...11S,2016ApJ...819...48S,grmonty}.

\subsubsection{Geodesic Calculation}
\label{sec:geodesic}
\label{sec:geodesic-calculation}

Because of the time-dependent spacetime, a unique geodesic must 
be calculated for  each pixel and snapshot.  The field of view
and the number of pixels determines the angular resolution of the
resultant images and the initial conditions of the geodesics.  The
quantity $\rmax$ sets the extent of the field of view, so the angular
field of view is therefore $\approx \rmax/\rcam$.  The simulation
domain is a sphere of radius $260M$; depending on how much of the
domain we want to ray-trace, we set $\rmax$ somewhere between $0M$ and
$260M$.

To mimic an observer at infinity, the camera must be sufficiently
distant for the spacetime to be nearly flat and for the rays shot out
to be nearly parallel in the regions of interest.  To achieve these
conditions, we place the camera at $\rcam = 1000M$ from the center-of-mass.
In order to confirm that this is an appropriate choice, we have
checked that $F_\nu(\rcam,\thcam,\phcam)$ changes by less than $1\%$ when moving
$\rcam$ from $10^3M$ to $10^5M$.  We chose the former setting for our
image generation because it is computationally less demanding---a serious concern when
processing $\mathcal{O}(10^4)$ snapshots, each with $\mathcal{O}(10^6)$
pixels.

We can adjust freely the polar (or inclination) angle $\thcam$ and the
azimuthal angle of the camera $\phcam$.  The inclination angle is of crucial
importance: viewing the SMBBH face-on ($\thcam = 0^\circ$) is
qualitatively different from viewing it edge-on ($\thcam = 90^\circ$).
As $\thcam \rightarrow 90^\circ$, images become more dependent on
$\phcam$ (or the phase of the orbit) because relativistic beaming and 
double-lens effects introduce strong azimuthal dependence.

We use a Lagrangian approach to integrate the geodesic equation:
\begin{equation}
\frac{d^2 x^\mu}{d\lambda^2} + \Gamma^\mu_{\alpha \beta} \frac{dx^\alpha}{d\lambda} \frac{dx^\beta}{d\lambda} = 0 \quad , 
\label{geodesic-eq}
\end{equation}
where $\Gamma^\mu_{\alpha \beta}$ are the Christoffel symbols, and 
$\lambda$ is the affine parameter. 
In practice, Equation~(\ref{geodesic-eq}) is cast in
$1^\mathrm{st}$-order form, providing us with a set of $8$ ODEs to
solve at each spacetime point for each ray, $4$ equations each for the
ray's position $x^\mu$ and 4-velocity $N^\mu$.  However, one equation
is eliminated because we find the time-component $N^t$ from the
null-like condition the ray must satisfy: $N^\mu N_\mu = 0$, choosing
the positive $N^t$ solution so that the ray points forward in time.
The initial direction of the spatial components, $N^i$, is chosen so that the
ray points toward the camera's focus at the position of the camera.

In this work, we use \bothros in the fast-light
approximation, which considers the simulation data frozen in time as
the photon travels through it.  The assumption simplifies the
simulation data-processing, as it allows us to ray-trace one time-slice
of data at a time.  However, we still include the time-dependence of
the metric by including its time derivatives in
Equation~(\ref{geodesic-eq}).  The validity of this approach will be
discussed in Section~\ref{sec:caveats_fastlight}. 

\begin{figure}[htb]
\center
\includegraphics[width=\columnwidth]{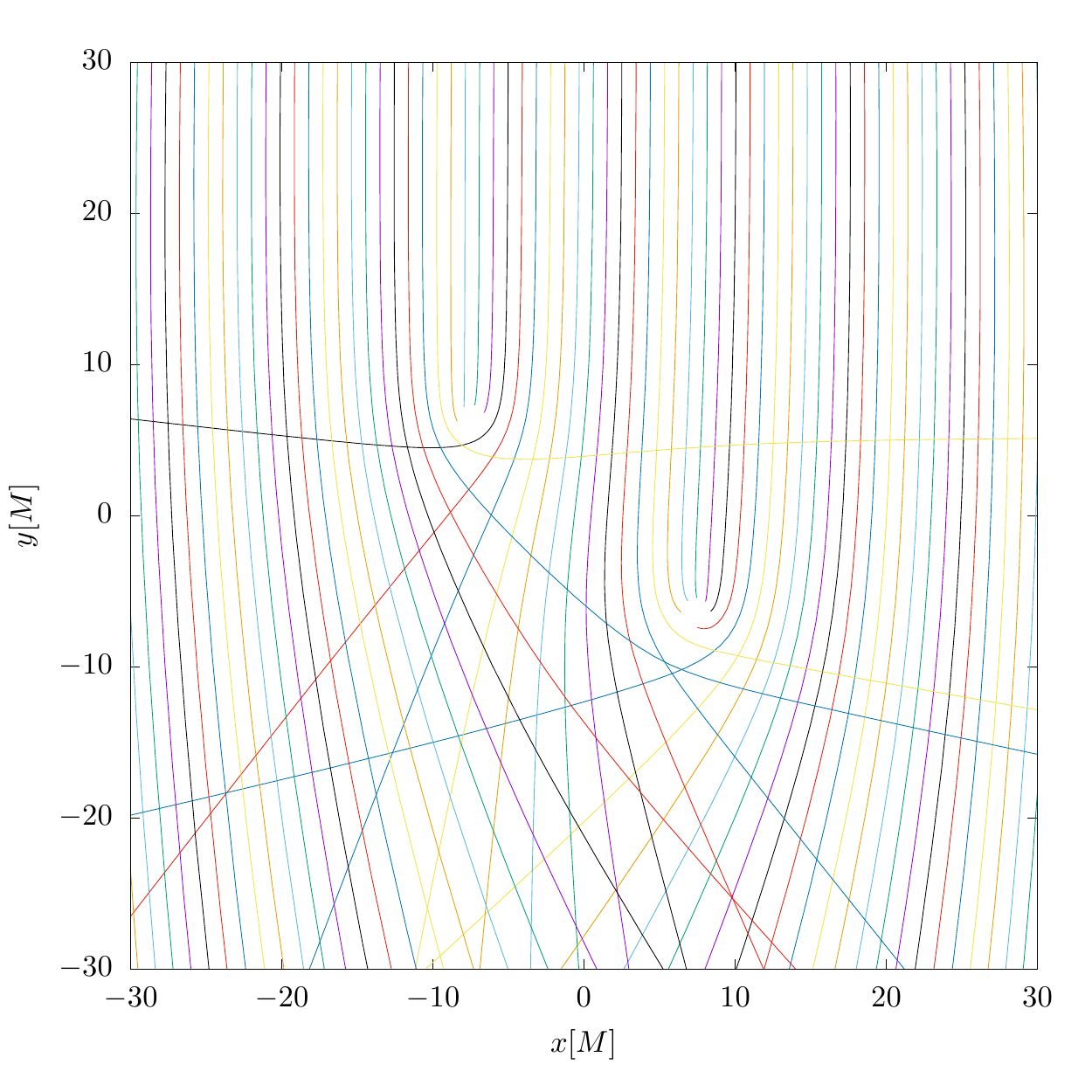}
\caption{Representation of 100 geodesics in the equatorial plane, distinguished using color. The $x$ and $y$ coordinates shown are in the PNH Cartesian coordinate system.}
\label{fig:geodesics}
\end{figure}

Because the camera is located at a distance $\rcam \gg a$ (where $a$
is the binary separation), photons must travel large distances of
empty (low emissivity and absorptance) space of little physical
interest. For computational efficiency and without loss of accuracy,
we may take larger steps through such regions than---for
example---near the BHs by exploiting an adaptive stepsize control
mechanism.  The $5^\mathrm{th}$-order Cash-Karp
algorithm~\citep{1992nrca.book.....P} is used because of its high-order
accuracy, stepsize adaptivity, and its ability to handle stiff
conditions.  We permit a maximum relative error of $10^{-6}$, which we
have demonstrated---through a convergence study---yields $\lt 1\%$
relative error in any quantities reported herein.
Figure~\ref{fig:geodesics} illustrates the results of our geodesic
calculations in the equatorial plane of the binary spacetime and gives
some visual intuition of the metric's influence on the rays.

\subsubsection{Processing the Simulation Data}
\label{sec:interpolation}

The cells in our simulation are
distributed nonuniformly in space using a time-dependent
transformation between a uniformly discretized numerical coordinate
system (${x^\prime}^\mu$)---analogous to a 3-d block of memory---and the actual spatial
coordinates of the cells expressed in a ``physical'' spherical
coordinate system.  The nonuniform system allows us to efficiently
resolve small features near the BHs as well as the larger scale
dynamics in the circumbinary disk.  We call this system ``warped'' or
dual-fisheye coordinates~\citep{WARPED}.  The simulation data are stored
on disks as a set of files, one file for each time slice or snapshot.
Each contains all the necessary 3-d grid functions $\left(\rho, \mathcal{L}_c, v^{\prime i} \right)$, where
$v^{\prime i}$ are the 3-velocity components of the gas in the simulation's coordinates.

Because the geodesic integration may be done independently from any
particular simulation's coordinate system and there are no symmetries
to exploit, we chose to generate the geodesics in the coordinates
used by our metric, namely the Cartesian PNH coordinate system
($x^\mu$).  The radiation transfer equation, however, depends on both the
geodesic information and the simulation data, so
processing~\harm simulation data requires us to transform
between $x^\mu$ and ${x^\prime}^\nu$, a problem requiring the solution of a nonlinear algebraic set of equations.
Grid function data
is interpolated to points along the ray, and vector quantities are transformed 
to the PNH Cartesian basis in which the geodesics are expressed.

For each point along a ray, the interpolation proceeds by first
converting the ray's coordinates into the simulation's coordinate
system, so that we may efficiently look up the subset of data needed.
Once the set of 8 cells surrounding a geodesic point are found, the
grid function values $\left(\rho, \mathcal{L}_c, v^{\prime i}\right)$
from these cells are linearly interpolated to this geodesic point.  To
match the simulation's resolution, we add a point (if necessary) on
the geodesic at each \harm cell encountered and interpolate in the
same way.  Sometimes the Cash-Karp algorithm used to integrate the
radiation transfer  equation requires values at
intermediate points along the ray; we compute these by
interpolating the stored values using the $4^\mathrm{th}$-order
Lagrangian method.  

The simulation's grid excludes ``cutout'' or ``excised'' regions
around the origin and $z$-axis in order to avoid coordinate
singularities.  Even though geodesics are free to move through these
cutout regions, no data is stored there.  This means that they do not
contribute to the ray's resultant flux or optical depth, so they are
treated as vacua for the sake of the transfer integration.

Once the GRMHD simulation data are read and interpolated, the
3-velocity ($v^{\prime i}$) of the gas is used to calculate its 4-velocity, $u^\mu$,
in global PNH inertial coordinates.  The fluid's 4-velocity 
allows us to calculate the fluid-frame frequency of the photon:
\begin{equation}
\nu = - k^\mu u_\mu , 
\end{equation}
where $k^\mu$ is the photon's 4-momentum.   Because the spacetime and the fluid's 
velocity field are inhomogeneous and dynamic, a photon's locally measured frequency can 
vary significantly along its path.

\subsubsection{Assigning Units}
\label{sec:units}

Lengths and times in the simulation are defined in units of $M$ (with $G=c=1$),
but because we performed the GRMHD simulation in the Cowling approximation 
(neglecting the fluid's self-gravity), the unit of gas mass is undefined. 
This fact allows us to set the total mass $M$ of the binary
and the physical density (or mass) scale of the gas independently 
when converting simulation data from code units to physical units.

To define a physical lengthscale appropriate to a SMBBH,
we set $M = 10^6 M_\odot$.  Instead of setting the gas density scale directly,
we derive it from a more intuitive quantity, the accretion rate $\dot{M}$.
To scale the accretion rate, we first calculate the accretion
rate in code units:
\begin{equation}
\dot{M}(r,t) = - \int \rho u^r  \sqrt{-g} \,d\theta d\phi ,
\label{mdot}
\end{equation}
where $u^r$ is the radial component of the 4-velocity, and the
integration is performed on spherical surfaces of fixed radius. In the
circumbinary region, these spheres are centered on the center-of-mass and the
radial component is that of the global inertial PNH basis.
Once $\dot{M}(r,t)$ in code units is found,
the density scale can be set by converting $\dot{M}$ to cgs units
as described in Appendix~\ref{app:conversion}.  Because the simulation did not achieve
inflow equilibrium in the mini-disks, while the inner portion of the circumbinary
disk in {\tt RunSS} did, we used the average $\dot{M}$
over the radial range $2a < r < 4a$ in our initial condition for
units definition.
We scale this value to a fraction of the Eddington accretion
rate in order to explore a range of optical thicknesses using the
parameter $\dot{m} \equiv \dot{M} / \dot{M}_\mathrm{Edd}$, where
$\dot{M}_\mathrm{Edd} = L_\mathrm{Edd} /(\eta c^2) $ with nominal
radiative efficiency $\eta = 0.1$ and $L_\mathrm{Edd} =
1.2 \times 10^{38} M / M_{\odot}~\mathrm{erg\ s^{-1}}$.
 We will start by studying high accretion rate flows ($\mdothigh$) in
Section~\ref{sec:supereddington}, then move to low accretion rate
flows ($\mdotlow$) in Section~\ref{sec:subeddington}.

The absence of inflow equilibrium in the mini-disks affects some of
our results; its implications will be discussed
in Section~\ref{sec:caveats_acc_rate}.   We completely ignore
emission from $r > 150M$, where the circumbinary disk is also out of inflow equilibrium,
but the neglected luminosity is much smaller.

In order to faithfully recover the flux of a simulated pointing, we need to ensure that
all parts of the simulation data are adequately sampled by the rays cast through it.  This means
that a given snapshot's flux, integrated over all the pixels, must be converged with respect
to the number of pixels used for a fixed field of view.  We have found that a resolution of
$\sim 7$~pixels/M  for the $r < 60M$ region and a coarser resolution
of $\sim 2$~pixels/M  for the $60M < r < 150M$ region is necessary to compute the
flux (at a given time and frequency) to $\sim 1\%$ accuracy.

\subsubsection{Radiation Transfer Solution}
\label{sec:rte}

The specific intensity measured at the camera is found by 
integrating the Lorentz-invariant form of the transfer equation along a geodesic:
\begin{equation}
\frac{\partial I}{\partial \lambda} = j - \alpha I  \quad ,
\label{rte}
\end{equation}
where $\lambda$ is the affine parameter, and $I$, $j$ and $\alpha$ are,
respectively, the Lorentz invariant intensity, emissivity and
absorption coefficient.  See Appendix~\ref{app:RTE} for details on the choice of
affine parameter and a derivation of this equation.  

The light coming from an optically thick medium is
effectively radiated from its photosphere (the surface at which its
optical depth passes through unity). If a geodesic encounters a
photosphere, we integrate the transfer equation from there to the camera with
initial condition $I_0 = I_{\mathrm{photosphere}}$; otherwise the
integration starts from the end of the geodesic with $I_0 = 0$. In
other words, we have to integrate the equation only in optically thin
regions.

In order to create spectra, we vary the frequency in cgs units at the 
camera, $\nu_{\infty}$. At each point $X$ along the
geodesic, we calculate the corresponding Doppler and gravitationally shifted
frequency $\nu_X$ via
\begin{equation}
\left.\nu\right|_X = - F(\nu_{\infty}) \left. \left(k^\mu u_\mu \right) \right|_X  \quad ,
\end{equation}
where $F(\nu_{\infty})$ is the conversion factor from numerical units
to cgs units (see Appendix~\ref{app:conversion}), 
$k^\mu$ is the 4-velocity
of the photon, and $u^\mu$ is the 4-velocity of the gas in global inertial
coordinates. Given a fluid-frame model for $j_\nu$ and $\alpha_\nu$,
we can then find $j(X)$ and $\alpha(X)$ from Equation~(\ref{inv-variables})
to discretely integrate Equation~(\ref{rte}).  The
integration is performed using the same $5^\mathrm{th}$-order
Cash-Karp algorithm used to integrate the geodesic equations.

\subsubsection{Opacity Model}
\label{sec:tau}

The radiative model chosen in a region strongly depends on whether the
gas is thermalized. Gas inside a disk can generally be considered
thermalized if the vertically integrated \textit{effective} optical
depth of the disk, $\tau_{\mathrm{eff}} \sim \sqrt{\tau_a(\tau_a
+ \tau_s)}$, is much larger than unity, where $\tau_a$ and $\tau_s$ are
the optical depths from absorptive and scattering processes,
respectively.  At the densities and temperatures of interest here, the
dominant source of opacity is electron scattering, so we neglect
absorptive processes.  Even though we do not calculate $\tau_a$ and,
rigorously speaking, $\tau = \tau_a + \tau_s \gg 1$ does not ensure
thermalization, we make the reasonable assumption that there is still
enough absorption in the disks for them to be thermalized.

We therefore assume a grey (frequency-independent) Thomson opacity for
electron scattering:
\begin{equation}
\alpha_\nu = \kappa_T\rho ,
\label{alpha}
\end{equation}
where $\rho$ is the gas density and $\kappa_T = \sigma_T / m_H$ is the
Thomson opacity (obtained by dividing the Thomson cross section by the
mass of an hydrogen atom).  Each segment along the ray contributes 
a Lorentz-invariant optical depth differential equal to
\begin{equation}
d\tau \ = \ \alpha_\nu \, ds \ = \ \alpha \, d\lambda \quad .
\label{tau}
\end{equation}
Here, we are interested in the optical depth between the observer and
the material we simulated, so we integrate Equation~(\ref{tau}) from the camera to the
source along the geodesic.

\begin{figure}[htb]
\center
\includegraphics[width=\columnwidth]{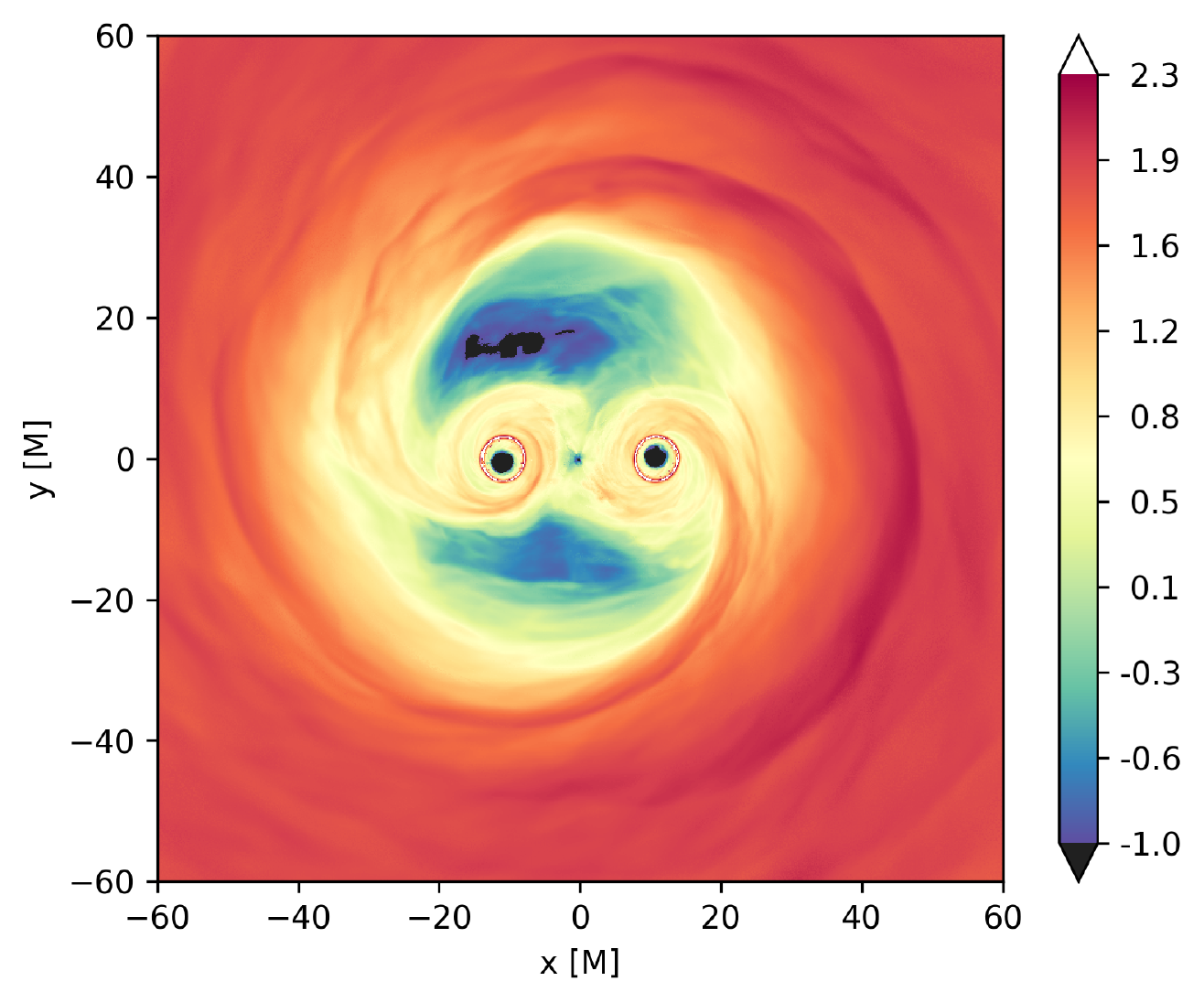}
\caption{Log$_{10}$ of the optical depth at $\mdothigh$, $\rmax = 60M$, $\thcam = 0^\circ$, and $t = 1180M \simeq 2 t_{\rm bin}$.
The rings of large optical depth circling the horizons correspond to the photon
spheres, where the geodesics wrap around the BHs multiple times,
accumulating extra optical depth.}
\label{fig:tau_00}
\end{figure}

In Figure~\ref{fig:tau_00}, we show the calculated optical depth in
the high accretion rate case at inclination $\thcam =
0^\circ$, which effectively corresponds to a vertical
integration. We see that the image bifurcates neatly into two zones:
those whose geodesics encounter so much gas (in mini-disks, accretion streams,
to the circumbinary disk) that $\tau \gg 1$, and those whose geodesics
traverse only cavities, so that $\tau \ll 1$ even after integrating to the
far end of the geodesic.

\subsubsection{Emissivity Model}
\label{sec:lum}
\begin{figure}[htb]
\center
\includegraphics[width=\columnwidth]{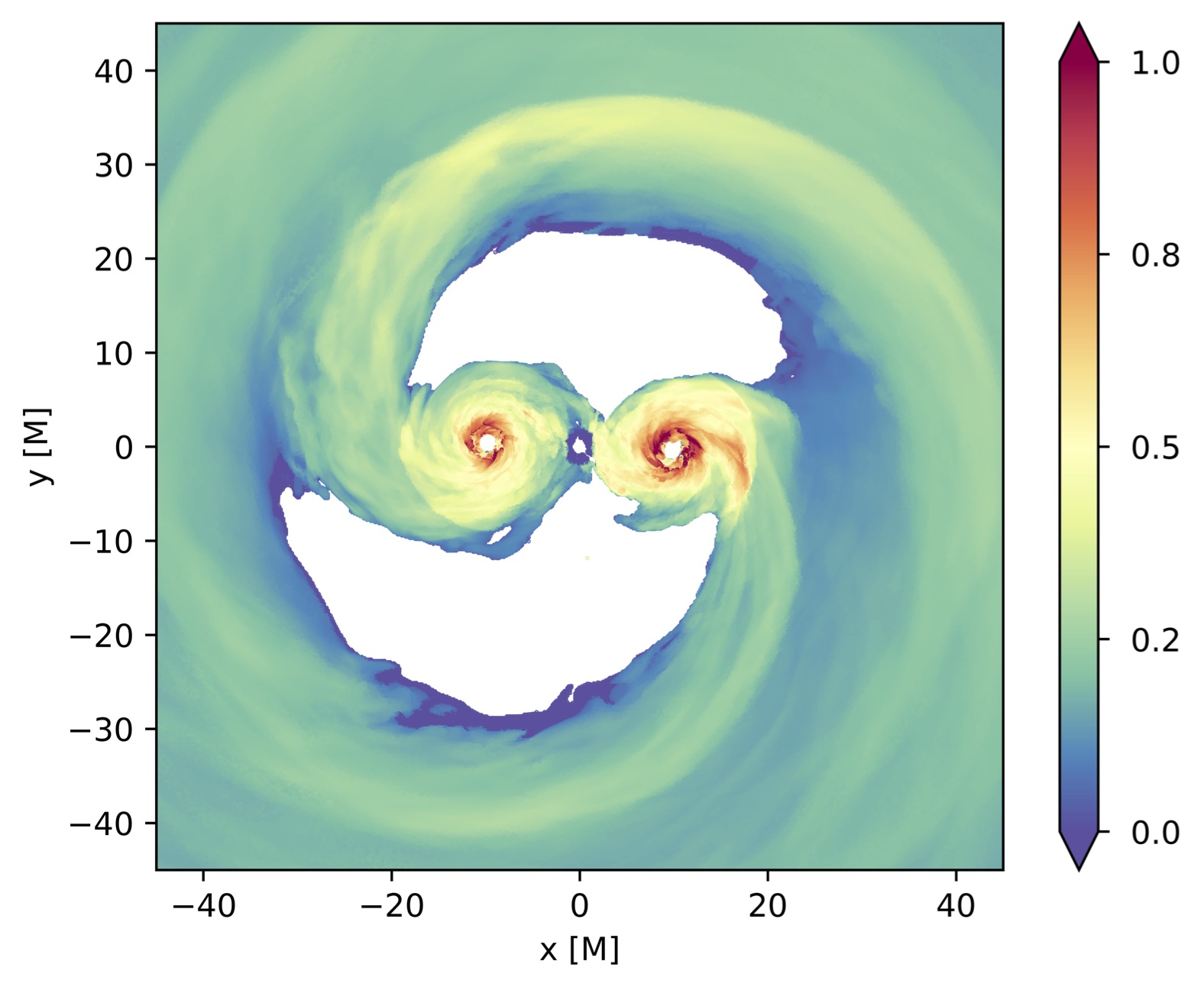}
\caption{Representation of the effective temperature on the photosphere,
projected into the binary's orbital plane, at $\mdothigh$, $t=1030M$.  Log$_{10} (\Teff/T_{0})$
in the fluid frame is shown, where $T_{0} = 5 \times 10^5$K.
The effective temperature at infinity is altered by gravitational redshift
and Doppler-boosting; the former dominates for face-on views, so the observed
effective temperature seen at $\thcam\simeq 0^\circ$ would be rather lower
near the BHs than shown here.  Uncolored (white) areas within the
cavity region lack the opacity necessary to surpass the photosphere criterion. }
\label{fig:photosphere}
\end{figure}

Below the photosphere (the $\tau = 1$ surface), we assume the disk's gas is in thermal equilibrium.
We therefore initialize the specific intensity at the photosphere with a black-body spectrum,
\begin{equation}
I_\nu = B_\nu(\nu,\Teff) = \frac{2h\nu^3}{c^2} \frac{1}{e^{\frac{h\nu}{k \Teff}}-1} .
\label{Inu-blackbody}
\end{equation}
The effective temperature, $\Teff$, 
is the temperature associated with the local
radiative cooling flux ($\mathcal{F}$) at the photosphere and can be found using 
Stefan-Boltzmann's law:
\begin{equation}
T_{\mathrm{eff}} = (\mathcal{F}/\sigma)^{1/4}  \quad , 
\end{equation} 
where $\sigma$ is the Stefan-Boltzmann constant. 
The flux is found by integrating the cooling function vertically
inside the photosphere:
\begin{equation}
\mathcal{F} = \frac{1}{2} \int _{\tau>1} \mathcal{L}_c \, ds 
			= \frac{1}{2} \int _{\tau>1} \mathcal{L}_c \, \nu \, d\lambda  \quad , 
\end{equation} 
where the factor of $1/2$ comes from the fact that the disk has two
photospheric surfaces from which heat can escape\footnote{Even
though we locate the $\tau=1$ surface as the photosphere, we ignore it if 
the total optical depth along the ray,
$\tau_\mathrm{tot}$, is $\tau_\mathrm{tot}<2$. This condition ensures that
the ``top'' photospheric surface  (i.e. the one
found by integrating through the disk from above) lies
above the ``bottom'' photospheric surface (i.e. the one found by
integrating through the disk from below).}.  This integral is approximately
vertical through the disk for views with $\thcam\simeq 0^\circ$; we
restrict our exploration of optically thick models to this viewing
angle because applying our ray-tracing method at other viewing angles
would locate the photosphere at an artificially high altitude from the
disk midplane.  Figure~\ref{fig:photosphere} gives a representation of
the photosphere and the effective temperature at its surface as viewed
face-on.

In evaluating this image, it should be recognized that in
those regions where the Thomson optical depth is $\sim 1$--10---most of
the mini-disks' surface area and part of the accretion streams'---a
substantial part of the dissipation may take place in regions that
actually lie outside the thermalization photosphere.  Consequently,
our approximation overestimates the thermal luminosity and underestimates
the luminosity arising from non-thermalized regions.

Outside thermalized regions, the predominant radiative process is
inverse Compton scattering.  In such regions, the dimensionless temperature
$\Theta \equiv kT/m_e c^2$ rarely exceeds $\simeq 0.2$ because the Compton
scattering energy-loss rate increases as electrons become relativistic and
further plasma cooling can be accomplished by pair production (see the pedagogical
review in \citet{krolik99}).
When the optical depth is small, the energy spectrum of photons is
exponentially cut off above $\Theta$ and follows a power-law below
$\Theta$.  In observed AGN, the power-law (for intensity) varies from
$\simeq -0.3$ to $\simeq -1.3$ \citep{Swift2017}; for simplicity, we
set it to -0.5.  We then take the emissivity to be
\begin{equation}
j_\nu \propto \mathcal{W}_\nu = \left(\frac{h\nu}{k T}\right)^{-1/2} e^{-\frac{h\nu}{k T}}  \label{jnu1}
\end{equation}
with $k T = 100 \kev$, or $\Theta \simeq 0.2$.  We normalize the spectrum in such a way that the bolometric
emissivity matches the cooling function at every point: $\int j_\nu d\nu d\Omega = \mathcal{L}_c$. 
This gives:
\beq{
j_\nu = \frac{\mathcal{L}_c}{4\pi A} \mathcal{W}_\nu , \label{jnu2}
}
\beq{
A = \int d\nu \mathcal{W}_\nu =  \frac{k T}{h} \sqrt{\pi} . 
\label{jnu-normalization}
} 
In these optically thin regions, the transfer equation is integrated starting with $I_\nu$ of the
disk at the photosphere (Equation~\ref{Inu-blackbody}) or zero if this geodesic
does not encounter a photosphere.  For $j_\nu$ and $\alpha_\nu$, it uses the emissivity
(Equation~\ref{jnu2}) and the scattering opacity (Equation~\ref{alpha}). 


\section{Results}
\label{sec:results}

As mentioned in Section~\ref{sec:units}, we chose a total system mass
$M=10^6 M_{\odot}$ to define the length scale and made two choices of
gas density scale through two choices of accretion rate in Eddington units:
$\mdothigh$ (high, Section~\ref{sec:supereddington}) and $\mdotlow$
(low, Section~\ref{sec:subeddington}).

\subsection{Optical Depth Images}
\label{sec:tau_snaps}

\begin{figure*}[htb]
  \centerline{\includegraphics[width=\textwidth]{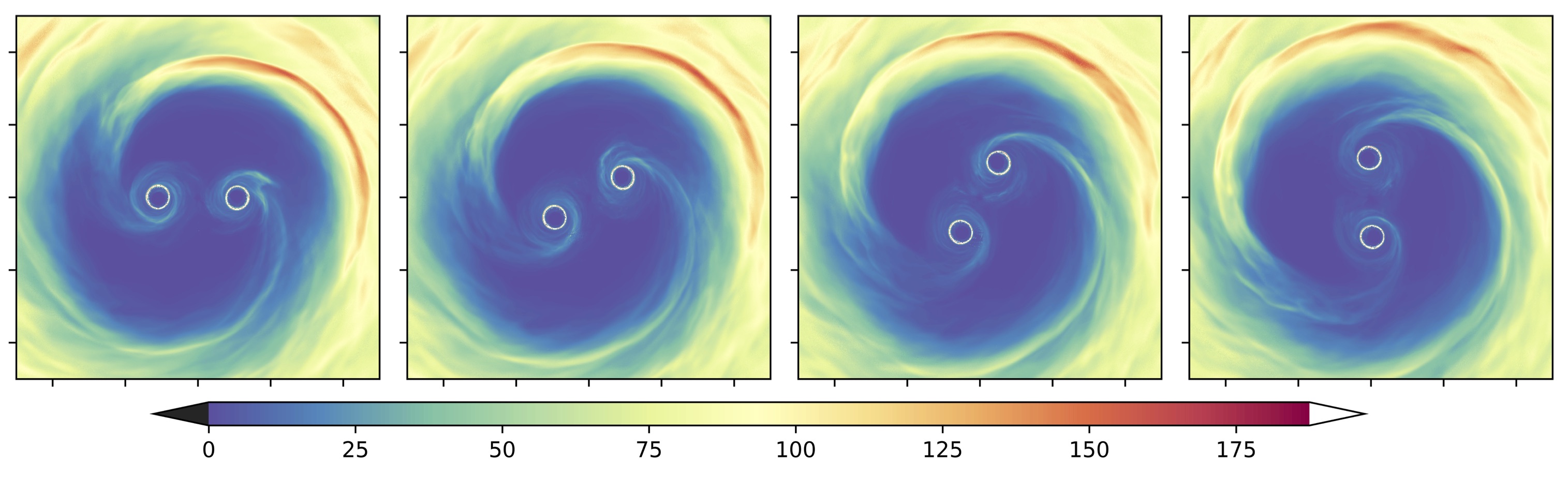}}
  \centerline{\includegraphics[width=\textwidth]{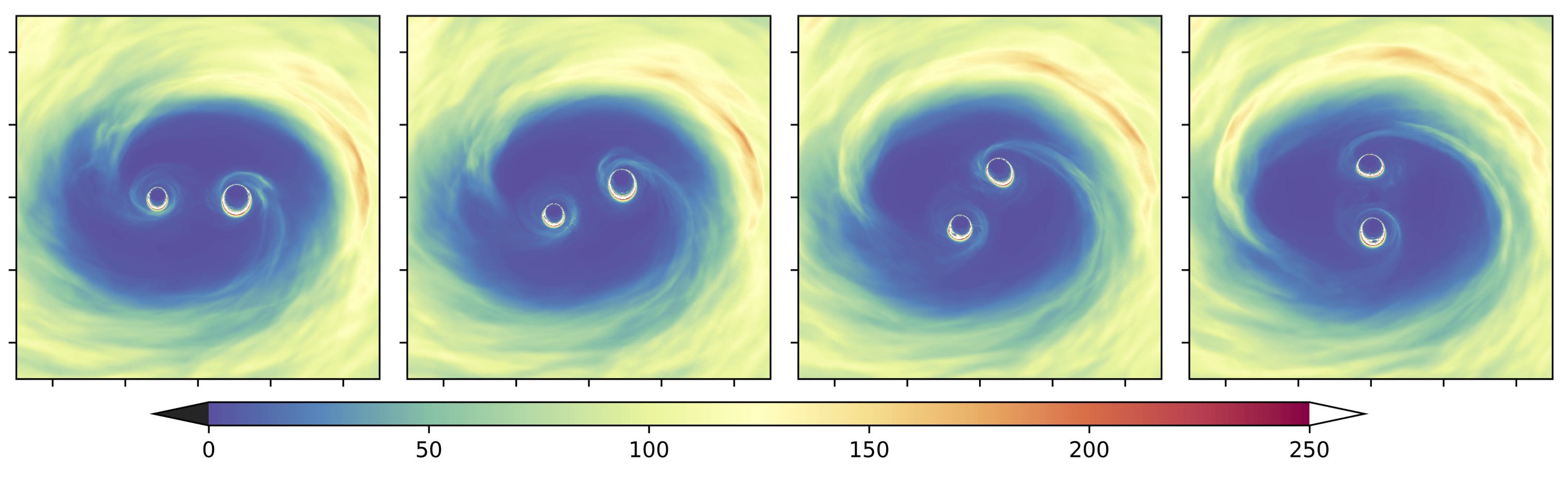}}
  \centerline{\includegraphics[width=\textwidth]{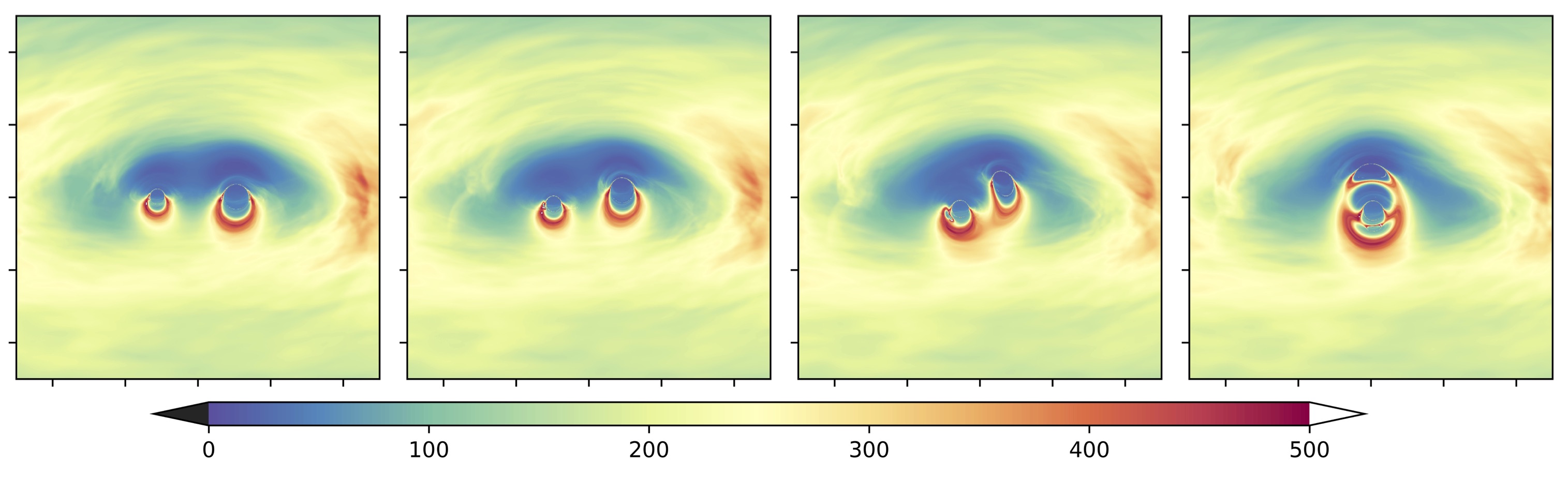}}
  \centerline{\includegraphics[width=\textwidth]{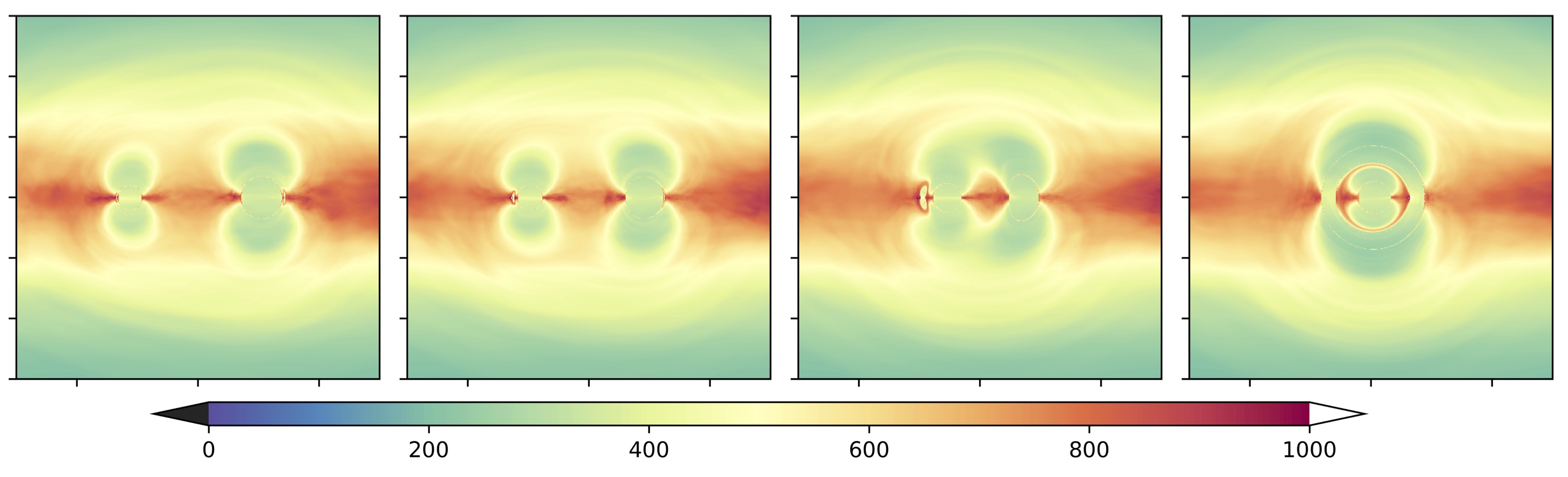}}
  \caption{Optical depth integrated along the line-of-sight at
  $\thcam = 0^\circ$, $39^\circ$, $71^\circ$, and $90^\circ$ from top to
  bottom, with $\mdothigh$.  Snapshot times are $1030M$,
  $1080M$, $1130M$, and $1180M$ from left to right. The outermost
  radius shown is $50M$, except at
  $\thcam = 90^\circ$, where it is set to $30M$ to
  focus on the complex relativistic effects at play. The space between tickmarks on all the axes is $20M$.}
  \label{fig:tau_snaps}
\end{figure*}

Thomson optical depth maps for $\mdothigh$ at a variety of polar
angles and times illustrate the basic
geometry of the system (Figure~\ref{fig:tau_snaps}); these images
can be readily scaled to other accretion rates because they are
linearly proportional to $\dot{m}$. 
The 16 panels show snapshots seen from four different polar angles at
four equally-spaced times spanning $150M$, a bit more than a quarter of a binary
orbit.

The face-on view ($0^\circ$ inclination) provides an approximate view
of the surface density of the gas.  The circumbinary disk is generally
quite optically thick ($\tau \gtrsim 50$), especially in the
overdensity or ``lump'' feature near its inner edge (\cite{Shi12,Noble12}).
As is usual for disks around binaries with order-unity mass-ratios,
the region within $\simeq 2a$ of the center-of-mass has very low
density except in a pair of spiral streams and in a pair of
mini-disks, one surrounding each member of the binary.  However, there
is also a high optical depth ring around each BH.  Rays reaching us
from this close to a BH wrap around it several times before heading
off to infinity, acquiring additional optical depth by traversing extra
path length. The characteristic magnitude of the optical depth
increases with inclination, reaching $\simeq 600$--900 at $\thcam =
90^\circ$, as the path through the BBH system is longer by a factor of
$\sin(\thcam)^{-1}$. Note, however, that the optical depth we measure at
edge-on views is not meaningful because real rays would traverse parts
of the disk well outside our simulation domain.

At intermediate viewing angles (e.g., the $39^\circ$ inclination shown
in the second row of Figure~\ref{fig:tau_snaps}), the optical depth
images still show the circumbinary disk geometry clearly. However, at
large viewing angles ($\ge 71^\circ$ inclination, the bottom two rows of
Figure~\ref{fig:tau_snaps}), gravitational effects distort the image
very strongly.  There is a region of high optical depth below the
BHs, where the photons travel through the circumbinary disk twice: 
starting above the circumbinary disk on the far side of the BHs, they pass downward
behind the BHs through the disk, curve through the cavity underneath,
and are finally gravitationally pulled upward (by the BHs) and traverse
a second time through the circumbinary disk toward the camera.
The low optical depth region above the BHs arises
from those photons which travel over the BHs and the circumbinary
disk and then curve upward through the cavity, avoiding the dense gas
found in the disks altogether.

Subtler relativistic features also appear at large viewing angles.
A mass moving along the line of sight to an observer creates a
gravitationally-lensed image of a source on its far side that is smaller
than its true size if the BH approaches the observer~\citep{Heyrovsky2005}.
More exotic but perhaps less apparent, the appearance of a
secondary image of one BH on the opposite side of the other BH can be
noticed; this image is due to the extreme light deflection close to the
horizons. A good example can be seen in the third frames of the bottom two rows of
Figure~\ref{fig:tau_snaps}), where a small oval feature forms on the left
side of the BH on the left. Regrettably, it is highly unlikely that
any of these imaging features will be spatially resolvable in the
foreseeable future.

\subsection{High Accretion Rate}
\label{sec:supereddington}

Our high accretion rate case, $\mdothigh$, is designed to
demonstrate the interplay between optically thick and optically thin
regions.  In the former, dissipated heat emerges in a thermalized
spectrum; in the latter, it is radiated by inverse Compton scattering
from a very hot electron population.  

Source-integrated features such as spectra may be observable soon.  To
compute spectra from our data, we take two time-averages of the flux from face-on
viewing angles, one over the second binary orbit, the other over the third.
The averaging suppresses statistical fluctuations.  We prevent blurring of
the image by rotating the camera orientation $\phcam$ at the binary orbital frequency.
We choose a face-on view at which our criterion for
distinguishing thermalized from coronal regions, namely whether the ray
optical depth is greater or less than unity, is well-justified.  At
higher inclinations it becomes increasingly suspect for two reasons. First, the optical depth unity point on a geodesic is found above the
actual photosphere because the path-length is $\propto \sec\thcam$.
Second, actual systems viewed edge-on may be obscured by material
at distances not included in our simulation.

\begin{figure*}[htb]
\center
\includegraphics[width=0.49\textwidth]{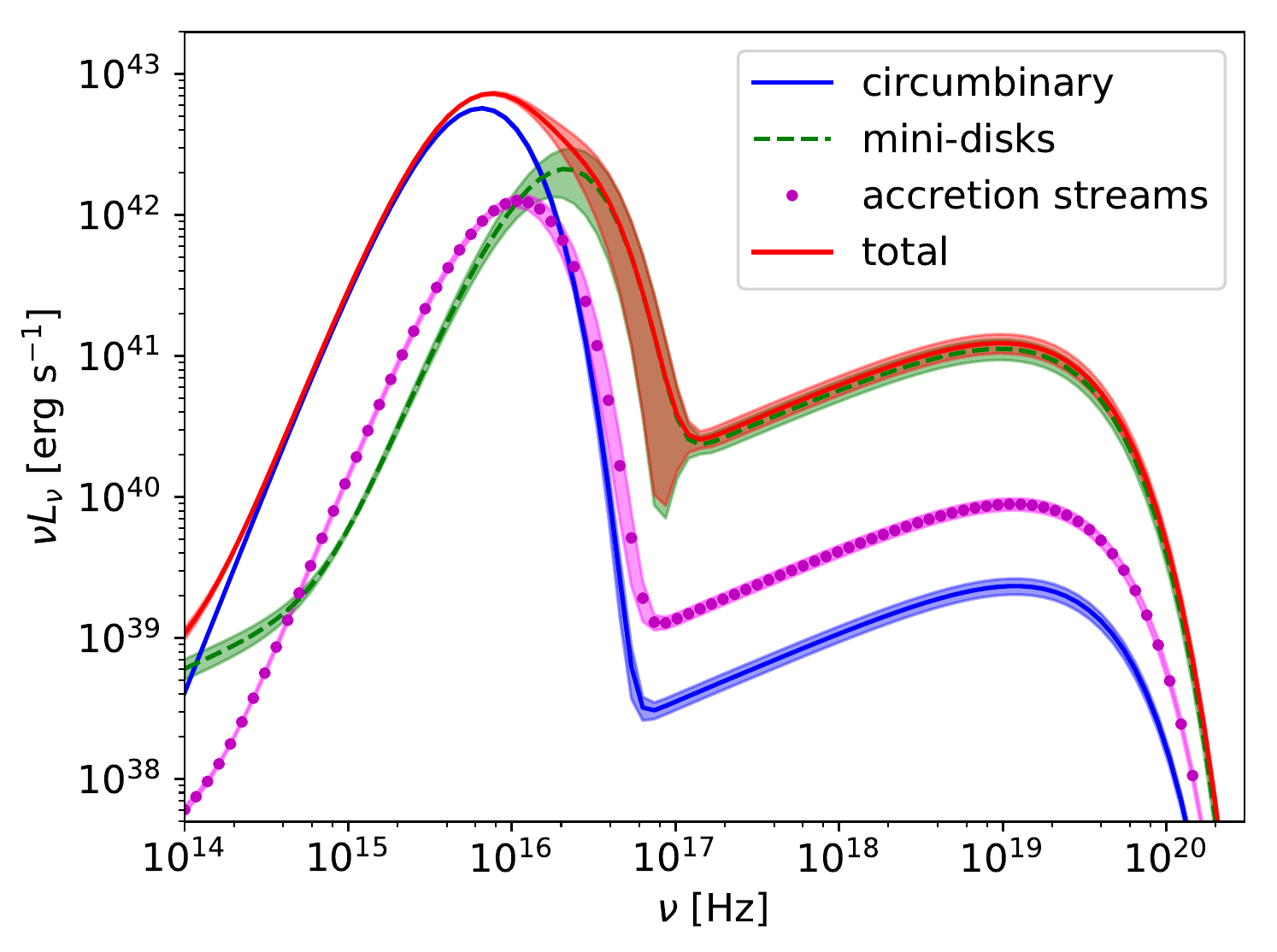}
\includegraphics[width=0.49\textwidth]{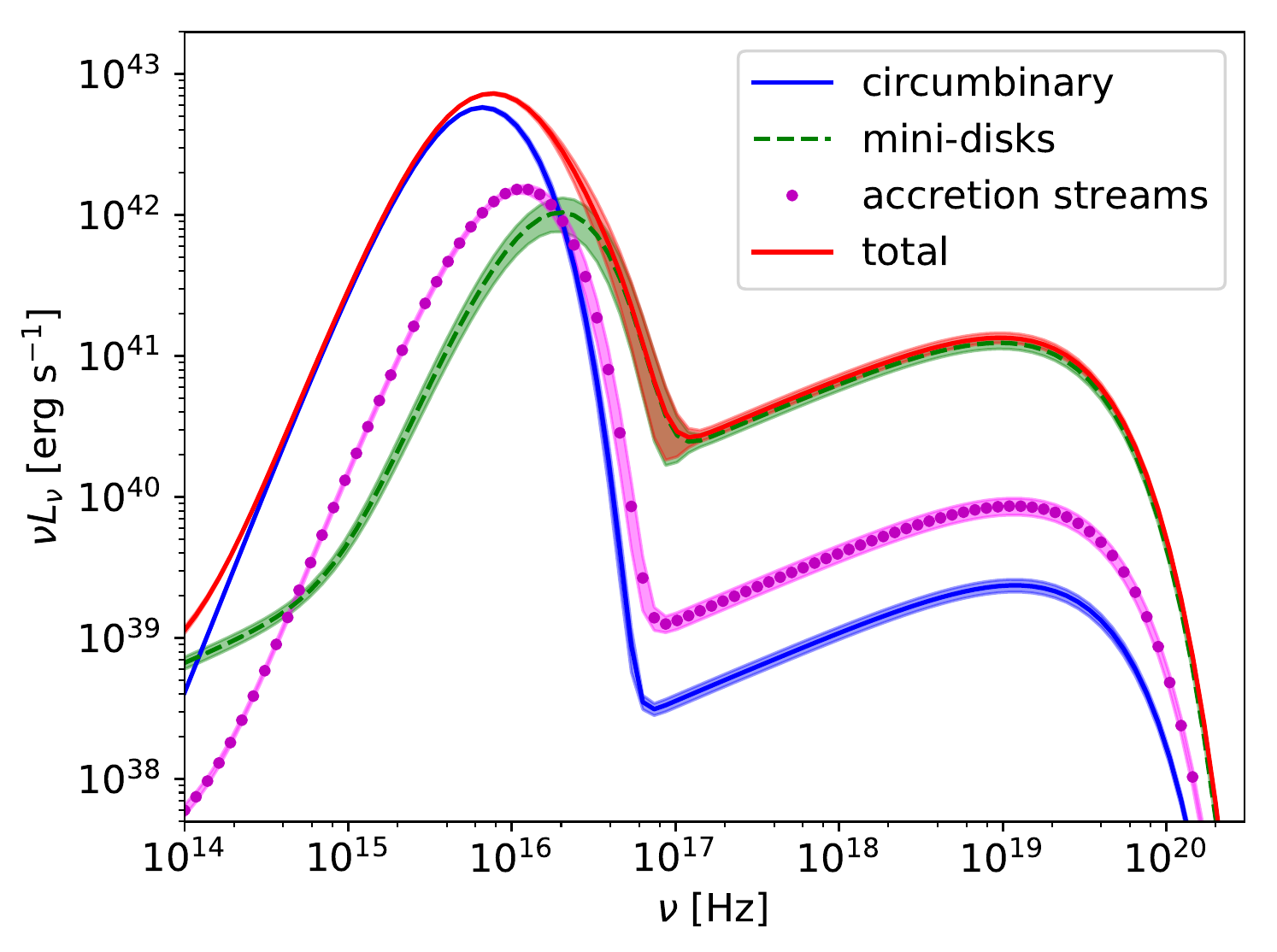}
\caption{Time-averaged luminosity ($\nu L_\nu$) spectrum obtained at 
$\thcam = 0^\circ$ and $\rcam =1000M$ with $\mdothigh$ using
simulation data from the second orbit (left) and the third orbit
(right).  We have separated contributions from the mini-disk regions
($r < a$), the accretion streams ($a < r < 2a$) and the circumbinary
region ($r > 2a$). The shaded region around each curve represents the
temporal variability of each component, one standard deviation above
and below, using a cadence of $10M$ for each orbit (60 samples).  The
cusps on the lower side of the mini-disks' shaded region
in the left-hand panel represent points lying off the scale because the
mini-disks' thermal emission fluctuates by an order of magnitude in
the second orbit. }
\label{fig:spec_thick}
\end{figure*}

\begin{figure*}[htb]
 \centerline{\includegraphics[width=\textwidth]{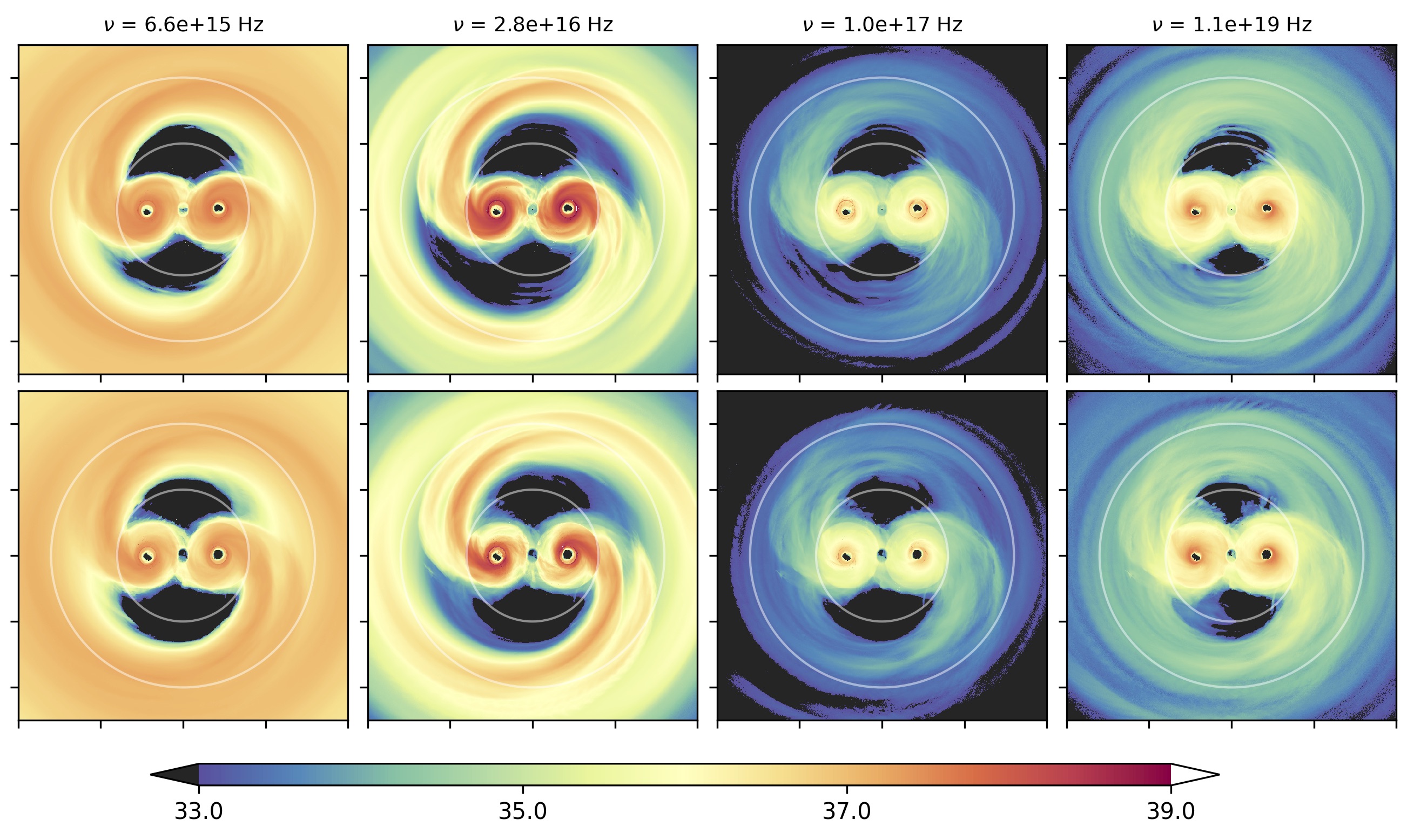}}
 \caption{
  Images of time-averaged spectral power ($4 \pi \rcam^2 \, \nu I_\nu $)
  [$\mathrm{erg}\ \mathrm{s}^{-1} $] at various
  frequencies with $\mdothigh$, showing the transition from the
  black-body dominated regime to the inverse-Compton scattering
  dominated regime. The time-averages were performed over the second
  orbit (top row) and the third orbit (bottom row) separately.  From
  left to right, we encounter (i) the circumbinary dominated UV, (ii)
  mini-disk dominated soft X-rays, (iii) X-rays near the boundary
  between thermal and corona dominance, and (iv) the mini-disk corona
  dominated hard X-rays.  The two white circles in each panel mark
  $r=a$ and $r=2a$. The space between tickmarks on all the axes is
  $20M$.}
\label{fig:spec_snaps}
\end{figure*}

Figure~\ref{fig:spec_thick} shows the spectral luminosity density in this
face-on view.  We define it as $L_\nu = 4\pi r_{\rm cam}^2 \int I_\nu \cos(\psi)d\Omega$,
where $\psi$ is the angle between the geodesic's direction at the
camera and the line of sight to the center-of-mass.  Images of the
system surface brightness at various frequencies in the face-on view are shown in
Figure~\ref{fig:spec_snaps}.

Like classical AGN spectra, this spectrum can also be described in
terms of two components: a thermal UV/soft X-ray portion and a coronal
hard X-ray spectrum.  The thermal UV originates from the photospheres
of the system; the hard X-rays are emitted in optically thin regions,
predominantly on the top and bottom surfaces of the disks.  All three
locales---the circumbinary disk, the accretion streams, and the
mini-disks---contribute to both the thermal and coronal spectral
components.  However, as previously remarked in Section~\ref{sec:lum},
our assignment of all dissipation below the unit optical depth surface
tends to transfer power from the Comptonized hard X-ray component to
the thermal component, and this effect is particularly strong in
the mini-disks and the accretion streams.

The emitted power is dominated by the thermal UV, with only $\sim 1\%$ 
radiated in hard X-rays; this ratio may, however, be exaggerated by our
simple emission model.  More surprisingly, the single greatest contribution
($\approx 65\%$) comes from the circumbinary disk rather than the mini-disks.  Because
we simulate a binary whose separation is only $20M$, the binding energy
of an orbit at $2a = 40M$ is 0.0125 in rest-mass terms; this is
more than half the effective radiative efficiency found in our simulation.
The mini-disks account for most of the remainder ($25\%$) in the second orbit,
but share the luminosity almost evenly with the accretion streams in
the third orbit.  The mini-disks are less luminous than would be expected
for the several reasons enumerated in the previous paragraph.

It is unsurprising that the thermal peaks from the three regions
should be found at frequencies that increase gradually from the
circumbinary disk to the accretion streams to the mini-disks.  In
time-steady ordinary accretion disks, the effective temperature is
$\propto r^{-3/4} R_R^{1/4}(r)$, where $R_R$ is a correction factor
accounting for the net angular momentum flux and relativistic
corrections.  This relation might be a reasonable approximation within
both the mini-disks and the circumbinary disk if $r$ is defined as the
distance to the near BH in the mini-disks and the distance to the
center-of-mass in the circumbinary disk.  However, the ``notch"
separating the circumbinary disk and mini-disk thermal spectra
predicted by \citet{Roedig:2014} is not apparent.  This can
likely be attributed to the comparative faintness of the mini-disks
in a system with binary separation as small as the one we have analyzed
(see Section~\ref{sec:spectra} for the arguments supporting this contention).

A more detailed analysis of where different frequencies are radiated
is aided by the images of Figure~\ref{fig:spec_snaps}.  The UV surface
brightness (first panel) in optically-thick regions varies hardly at all
from the circumbinary disk to the accretion streams to the mini-disks,
but the larger area of the circumbinary disk makes it
the primary contributor to the luminosity in this band. However, because the
mini-disks are warmer than the circumbinary disk, their thermal spectrum
remains bright farther into the EUV (second panel).  This image further
reveals that, especially in the third orbit, a sizable part of the dissipation
occurring in the mini-disks takes place in spiral shocks.
In the soft X-ray band (third panel), the principal contributor is the
extreme Wien tail of the thermal emission from the mini-disks.
Finally, in the hard X-ray band (fourth panel), the emission is dominated by
the optically thin component in the corona, which is strongly concentrated
in the innermost rings of the mini-disks.  Again, we caution that
the mini-disk radiation produced in our model may overestimate the
thermal component at the expense of the Comptonized X-rays.

The first two of the panels in Figure~\ref{fig:spec_snaps}
also show that nearly all the light attributed to the
accretion streams in Figure~\ref{fig:spec_thick} is associated with
the shock that occurs when the accretion stream, having
been strongly torqued by the binary's gravity, is flung outward
and strikes the inner edge of the circumbinary disk.  

A number of these comments are in agreement with the
2-d ``$\alpha$-viscosity" hydrodynamics simulations of \citet{Farris15b}.
They, too, found enhanced emission due to shocks between the accretion
streams and both the inner edge of the circumbinary disk and the
outer edges of the mini-disks.  In addition, because they assumed
all radiation was thermal, they placed all this light in the UV/EUV.
However, they also found the unshocked streams had high surface brightness,
a result attributable to the ``$\alpha$-viscosity" creating dissipation
even in laminar regions if they contained significant shear.  In
addition, their separation of ``mini-disks" from ``cavity" from ``circumbinary
disk" is different from ours, so the separate luminosity contributions
cannot be directly compared.   Although the images of surface
brightness are qualitatively similar to those of \citep{Tang:2018rfm}
at times well before merger in their simulations, and the mass of
the system they simulate is only twice that of ours, the effective temperatures
\cite{Tang:2018rfm} find ($\sim 1$~keV in the circumbinary disk, tens of keV in the mini-disks)
are much higher than ours.  It is possible this contrast arises because
their simulation treats a case with an accretion rate $\gtrsim 10^4 \times$ ours
(in order to support temperatures $\gtrsim 10\times$ higher on a similar
thermally-radiating surface area), but they
explicitly state neither the accretion rate they found nor the magnitude
of their initial gas surface density.

\subsection{Low Accretion Rate}
\label{sec:subeddington}
\label{sec:anisotropy}
As explained earlier, we have complete freedom to choose the
inclination of the camera only if the whole domain is optically
thin. To enable further study of inclination effects, we also studied
a low accretion rate, $\mdotlow$. Because the density
is $625\times$ lower in this case than in the high accretion-rate case, 
the system stays optically thin along all directions except those within
$\approx 20^\circ$ of the orbital plane. For high inclination angles
($\thcam \gtrsim 45^\circ$), the camera's azimuthal angle also becomes
important. Note that here the azimuthal angle is measured in the corotating
frame of the binary, with $\phcam = 90^{\mathrm{\circ}}$ when the camera
is aligned with the two BHs. If, as is more likely in reality, the
camera is stationary in an inertial frame, the camera's effective
azimuthal angle varies periodically as the binary members orbit.

\begin{figure*}[htb]
  \centerline{\includegraphics[width=\textwidth]{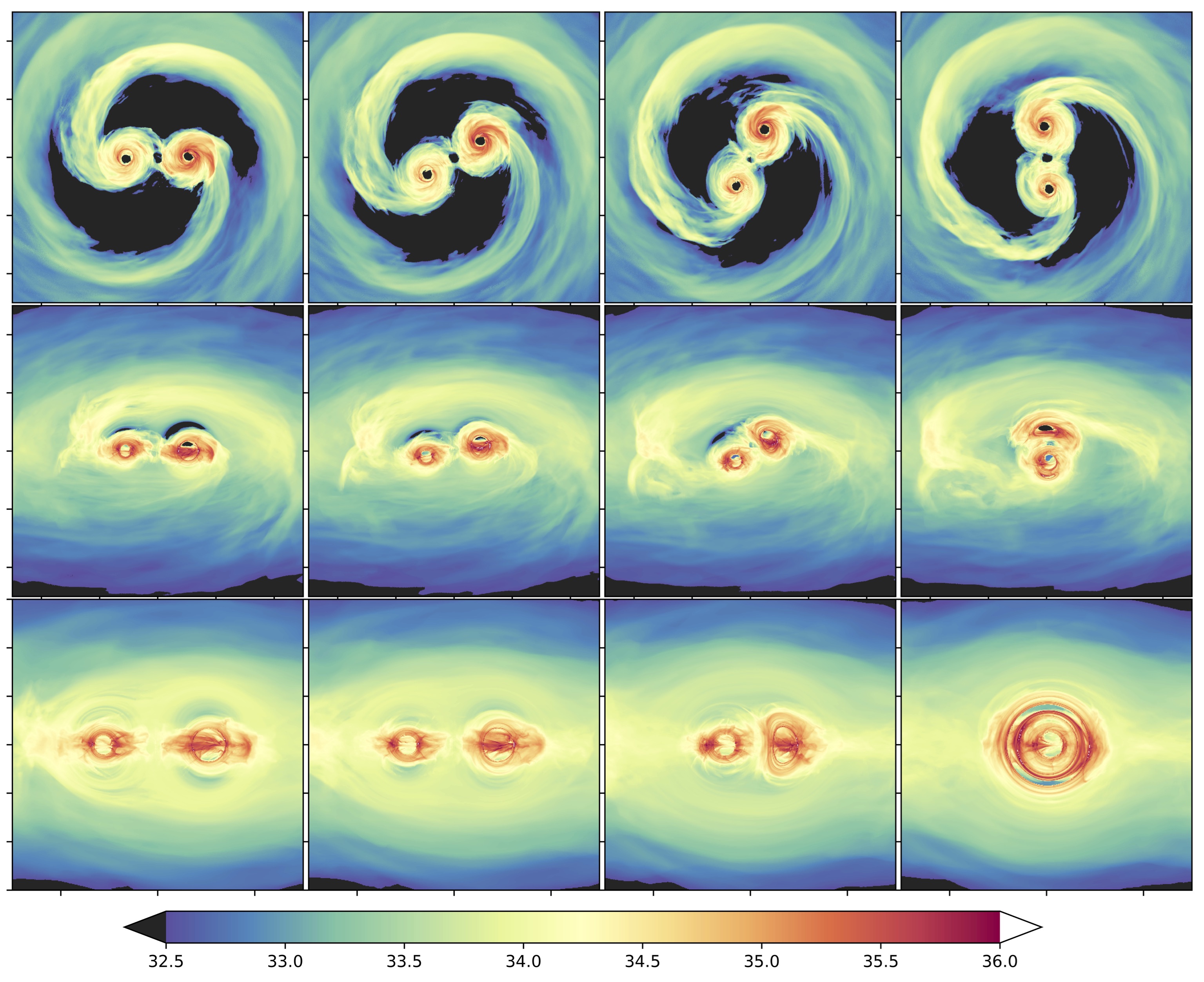}}
  \caption{Log$_{10}$ of the Spectral
  power ($4 \pi \rcam^2 \, \nu I_\nu $)  [$\mathrm{erg}\ \mathrm{s}^{-1} $]
  at $\nu = 10^{19}~\mathrm{Hz}$, $\thcam =
  0^\circ,~71^\circ,~90^\circ$ (from top to bottom) and
  $t=1030M,~1080M,~1130M,~1180M$ (from left to right), with $\mdotlow$.
  Again, the width of each image is $50M$, except at
  $\thcam = 90^\circ$ where it is set to $30M$ to
  give a better view of the gravitational lensing. The space between tickmarks on all the axes is $20M$.}
\label{fig:lum_snaps}
\end{figure*}

We first display images
of very hard X-ray ($10^{19}$~Hz, $\simeq 400$~keV) intensity at three
different inclination angles,
each viewed at four different times spread over a quarter-period
(Figure~\ref{fig:lum_snaps}).
These images somewhat resemble those of
optical depth presented in Section~\ref{sec:tau_snaps} because the
cooling function, like the scattering coefficient, is proportional to
the density.  However, while the optical depth of the inner circumbinary
disk was generally $\simeq 5\times$ the optical depth of the mini-disks,
the X-ray surface brightness of the mini-disks is $\simeq 5\times$
that of the circumbinary disk.  This is, of course, because the rate of energy
dissipation per unit mass is much higher so close to the BHs.

The spectrum of the optically-thin case looks very much like the
high-energy ($\nu > 3 \times 10^{16}$~Hz) portion of the spectrum
shown in Figure~\ref{fig:spec_thick} because we assumed that the
fluid-frame coronal spectrum is the same everywhere.  Relativistic
effects, however, lead to significant viewing-angle dependence.

\begin{figure*}[htb]
\centerline{
\includegraphics[width=0.49\textwidth]{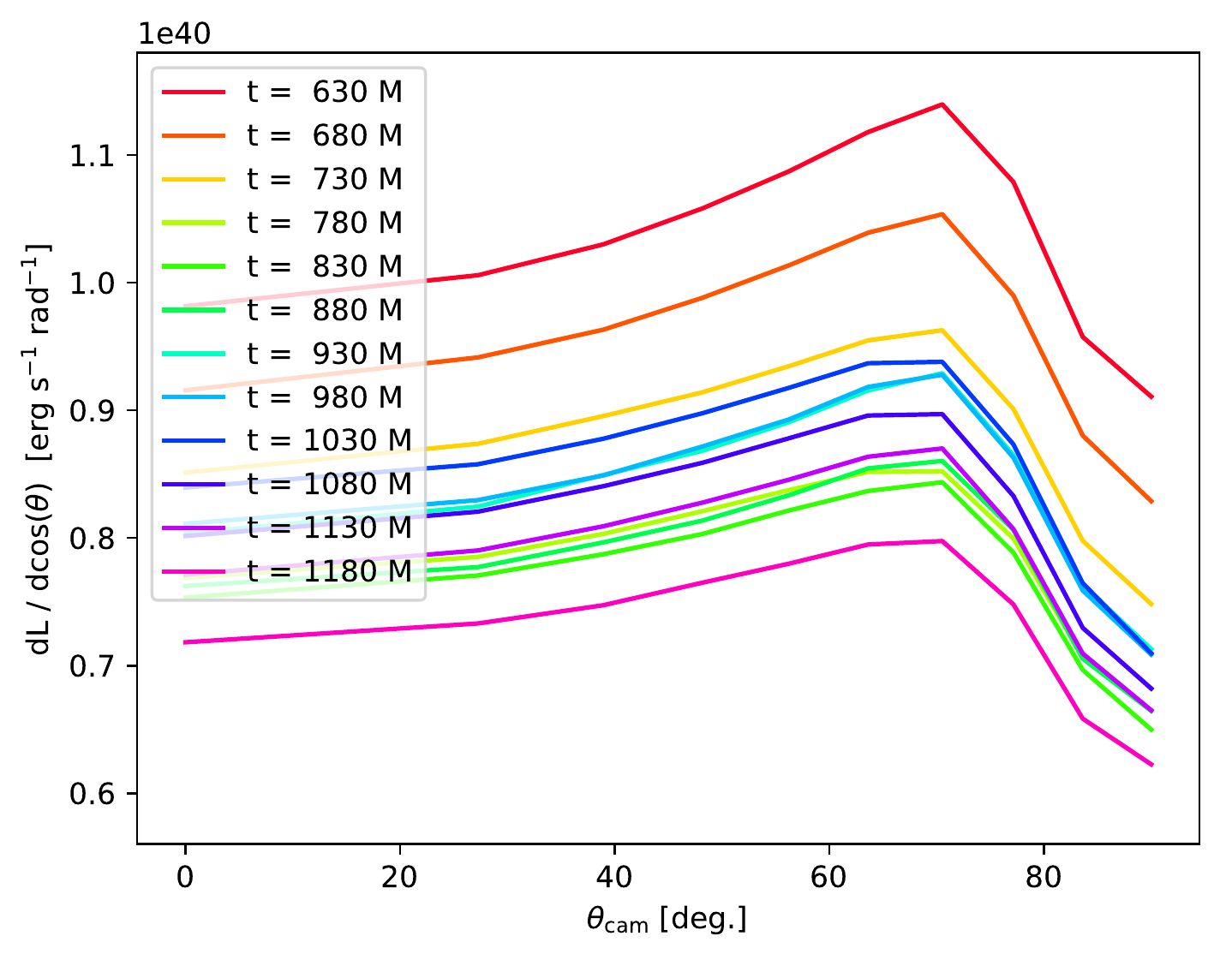}
\includegraphics[width=0.49\textwidth]{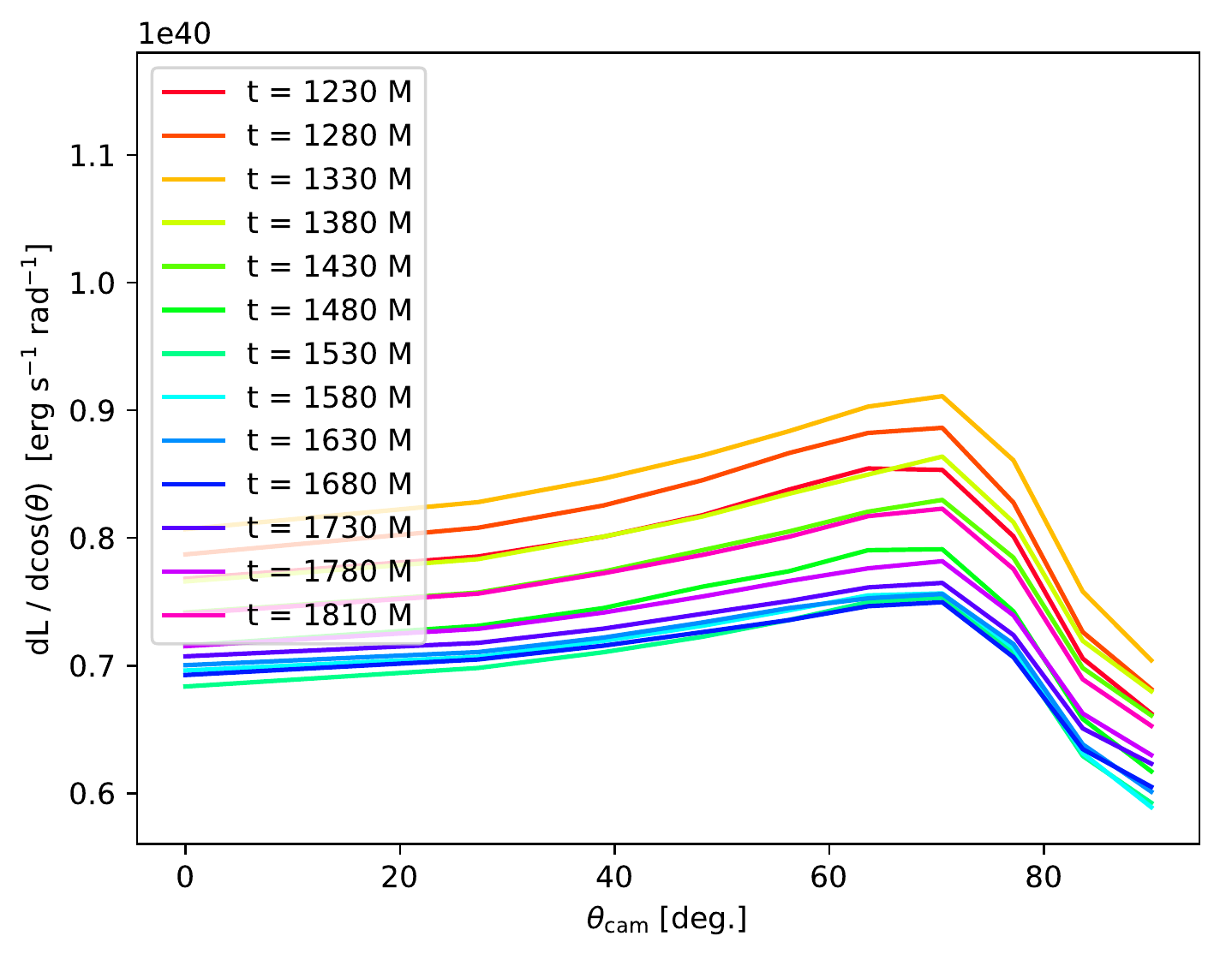}
}
\caption{Dependence of the bolometric luminosity on the observer's angle of inclination ($\thcam$) from the $+z$-axis for $\mdotlow$.
The luminosity was averaged over azimuthal viewing angle ($\phcam$), and time-averaged over the second orbit (left) and the third orbit (right). Curves at 12 equal intervals of time 
during each orbit are shown. }
\label{fig:lum_ang}
\end{figure*}
\begin{figure*}[htb]
\centerline{
\includegraphics[width=0.49\textwidth]{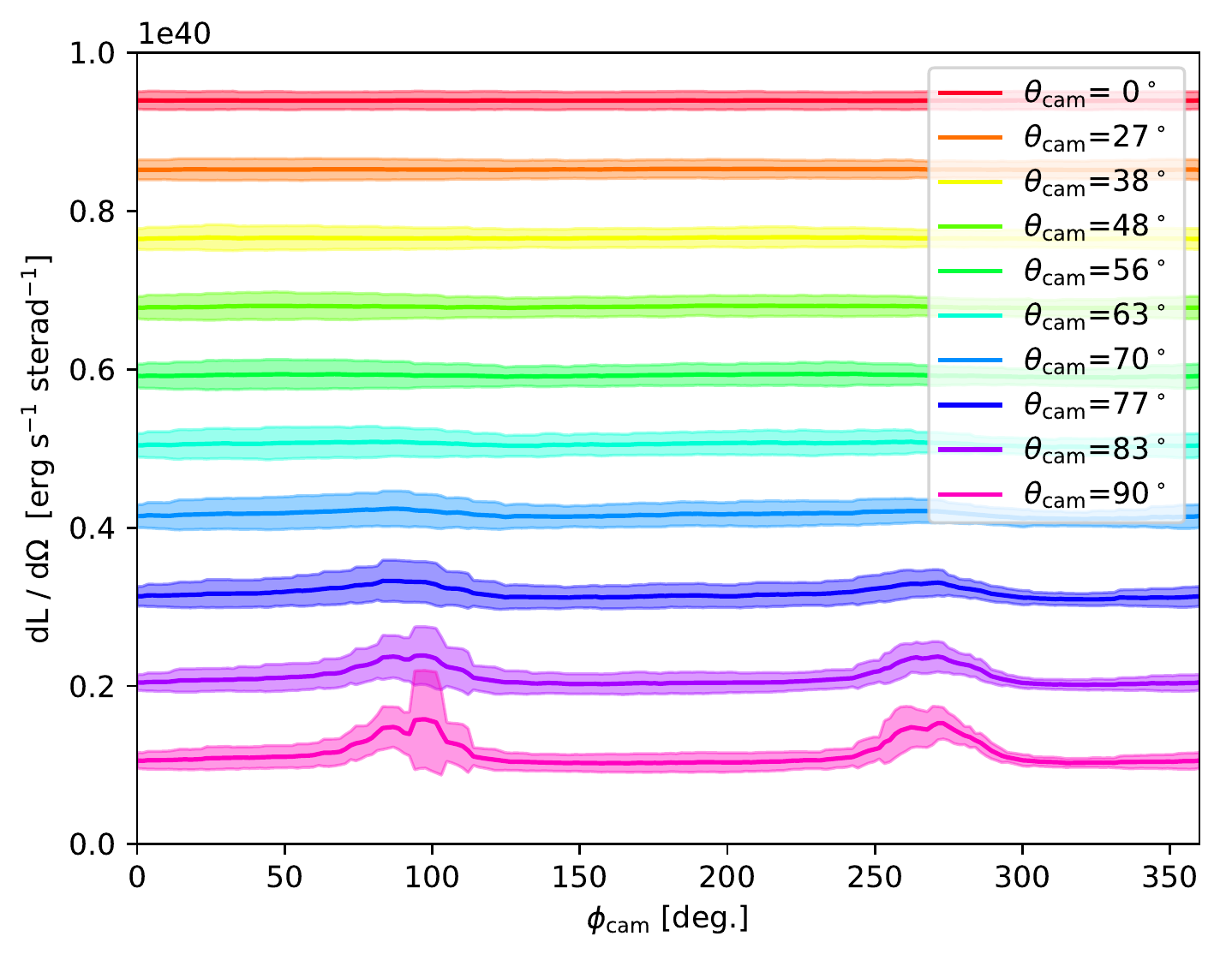}
\includegraphics[width=0.49\textwidth]{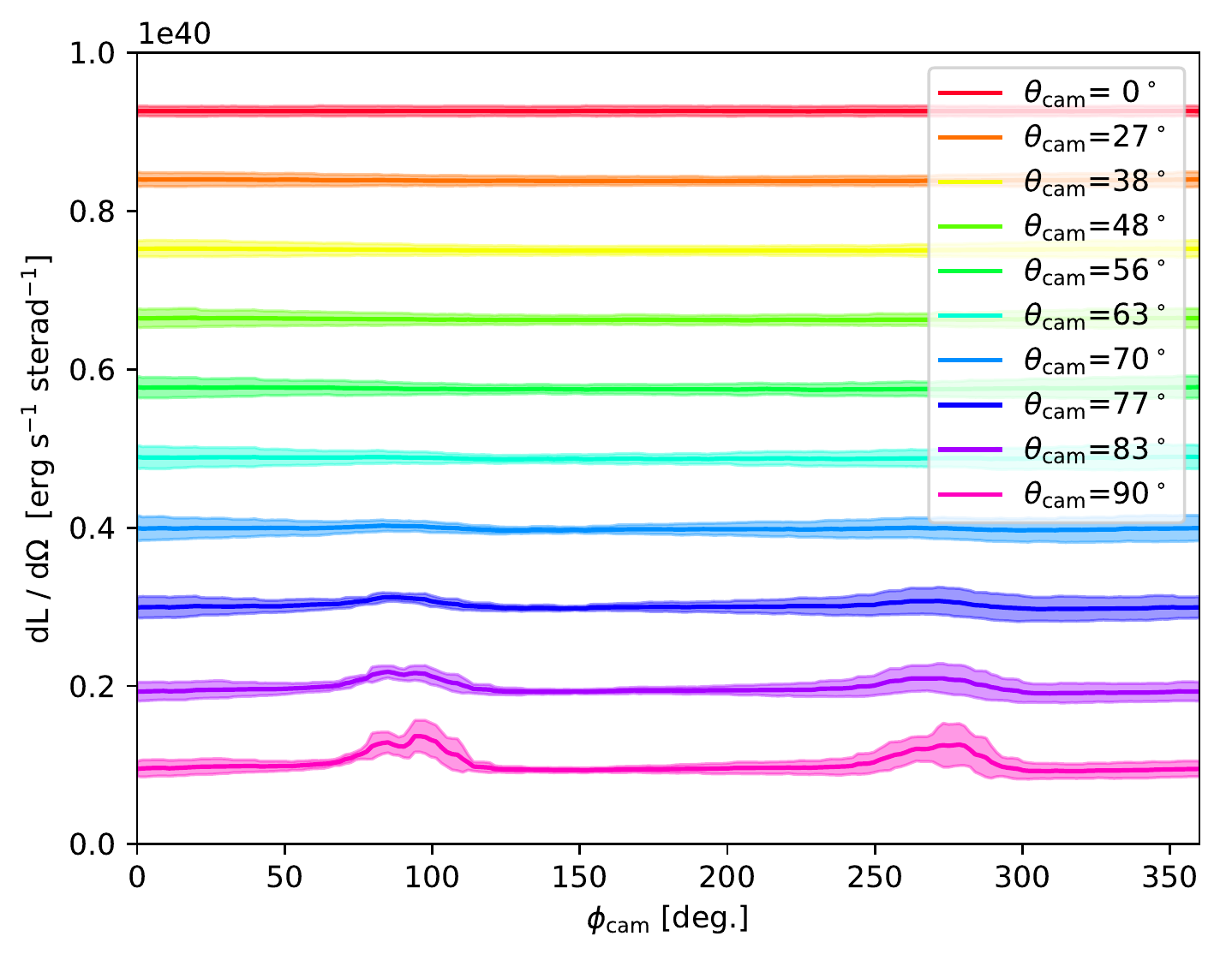}
}
\caption{Dependence of the bolometric luminosity on the
observer's azimuthal position ($\phcam$) at a variety of inclination
angles ($\thcam$) for $\mdotlow$.  Solid curves represent
time-averaged quantities over the second orbit (left) and the third
orbit (right), while shaded regions show the standard deviation of the set of 
12 time levels of data used to make the averages.  The $i^\mathrm{th}$
curve from the bottom is vertically offset by
$ 9 \left(i-1\right) \times10^{38} \mathrm{erg} \mathrm{s}^{-1} \mathrm{sterad}^{-1}$,
meaning the $\thcam = 0^\circ$ curve should really lie at
$\sim1.1\times10^{39} \mathrm{erg} \mathrm{s}^{-1} \mathrm{sterad}^{-1}$,
}
\label{fig:lum_sphere}
\end{figure*}

Although the shape of the spectrum depends only very weakly on viewing
angle, relativistic effects alter the angular distribution of its
intensity strongly enough to be observationally
interesting.  Figures~\ref{fig:lum_ang} and \ref{fig:lum_sphere} display
how the observed bolometric flux depends on both the polar and
azimuthal angle of an observer.
We show only bolometric luminosity because the spectral shape is
much less sensitive to viewing angle than its overall level.

At most inclinations ($\thcam \lesssim 70^\circ$), the flux is almost
independent of azimuthal angle, but increases with inclination due to
relativistic beaming of light emitted by gas moving toward
the observer.  However, for nearly edge-on viewing angles
($\thcam \gtrsim 70^\circ$), the angular-dependence becomes more complex due to
three effects. First, lensing of the farther BH by the nearer one
when $\phcam \simeq 90^{\circ}$ or $270^{\circ}$ significantly brightens the image;
the peak flux can be a factor of 3 higher than at other azimuthal angles with the same
inclination.
Second, relativistic beaming increases sharply with
greater polar angle.  Third, even for $\dot{m}$ as low as $\mdotlowvalue$,
the circumbinary disk can intercept light along rays
passing close to the orbital plane.  Smaller values of $\dot{m}$
would diminish the range of angles around the plane affected by
optical depth.  Because the bright peak is due to lensing, it
is not particularly affected by the optical depth through the
disk.  However, at other azimuthal angles, the disk cuts off
the rise in flux due to Doppler beaming,
running almost all the way around the orbital plane in which
the flux at $\thcam = 90^\circ$ is almost a factor of 2 smaller than
that at $\thcam = 70^\circ$.

\section{Discussion}
\label{sec:discussion}


\subsection{Spectral Features}
\label{sec:spectra} 

There has been much discussion in the literature regarding the imprint on the
thermal spectrum that may be created by the gap formed around a binary, with
suggestions ranging from a sharp cut-off at the temperature of the
circumbinary disk's inner edge (in early work assuming there is little
accretion from the circumbinary disk to the binary:
\cite{Tanaka:2012,Gultekin:2012,Kocsis:2012,TanakaH:2013}),
to a deep notch between frequencies corresponding to the temperature of
the circumbinary disk's inner edge and those corresponding to the temperature of the
mini-disks' outer edge \citep{Roedig:2014}, to a gentle change of
slope in this region \citep{Farris15b} or a distinct notch, but centered
at several keV \citep{Tang:2018rfm}. The degree to which
such a feature appears hinges on the contrast between the highest
temperature achieved in the circumbinary disk and the lowest temperature
found in the mini-disks; more precisely, the relevant contrast is between
the temperatures of regions radiating thermally.  Applying simple equilibrium Newtonian
disk theory to this situation, as in \citet{Roedig:2014}, leads
to a temperature ratio across the gap of $\simeq 3$.

In our high accretion rate example, this ratio is smaller, only $\simeq 1.25$--2
(see Figure~\ref{fig:photosphere}), smoothing the spectrum so that the notch
almost disappears (see Figure~\ref{fig:spec_snaps}).
The diminished temperature ratio results from a combination of effects,
none of them present in simple disk theory.

One is a higher temperature strip along
the inner edge of the circumbinary disk (also visible in Figure~\ref{fig:photosphere})
due to the shock driven by accretion streams flung outward by the binary's torque.  Also found in the work of \citet{Farris15b},
this strip can partially fill the notch if it radiates thermally.

The small binary separation we have studied ($a = 20M$) also leads to a weaker temperature
contrast: the outer edges of these mini-disks are only a factor of 2 outside their innermost stable circular orbit (ISCO) radii.
So close to the ISCO, the dissipation rate per unit area rises inward considerably
more gradually than the classical $r^{-3}$ scaling.   Incorporating this correction
into standard disk theory for spin-less BHs (as in our simulations) diminishes
the predicted temperature ratio from $\simeq 3$ to $\simeq 2$, but if the BHs had
near-maximal spin, there would be essentially no alteration to the temperature ratio
because the ISCO angular momentum is rather smaller for rapidly-spinning BHs
than for non-spinning BHs, and this flattening of the dissipation profile is
due to diminution of radial contrast in angular momentum. For essentially the same reason,
at radii only a few times that of the ISCO, radial pressure gradients can accelerate
inflow without dissipation \citep{IB01,KHH05}.  Thus, the surface brightness not
far outside the ISCO can be significantly depressed relative to the
classical $r^{-3}$ scaling. In addition, the short inflow time in the mini-disks when the binary 
separation is small leads to a state in which they are close to inflow equilibrium with 
respect to their instantaneous mass accretion rates, but this rate can be either 
larger or smaller than their mean share of the accretion rate through the circumbinary disk.

Thus, in the circumstances posited here, the notch is likely to be weak.
However, it could partly re-emerge in binaries with separations larger
by a factor of 2--3 or more or in binaries in which the BHs spin
rapidly because in both instances the outer rim of the mini-disks is farther
from the ISCO.  It might be further re-excavated if more
of what is considered ``disk" in our analysis were, in fact, optically
thin enough to make thermalization of its emission imperfect.
Even at an accretion rate $\dot m = 0.5$, the Thomson optical depth
through the outer rim of the mini-disks, where the accretion streams
strike, is only $\simeq 5$--10.  Particularly at the very high post-shock
temperature associated with these shocks, the absorption opacity is
likely too low to thermalize the radiation spectrum.  As a result,
energy would be shifted from radiation at the lowest-temperature of
the mini-disks to much harder photons, while the higher-temperature
thermally-radiating regions of the mini-disks would be unaffected.

Hard X-ray production is generic to accreting BHs, and is usually
attributed to inverse Compton scattering by electrons heated by magnetic
dissipation events in low density material above the accretion disk
\citep{SchnittmanKN:2013}.  At low accretion rates, the optical depth
of the mini-disks, the accretion streams, and the inner regions of the
circumbinary disk would be so small that essentially all the radiated
power should be in this band.   Post-processing of simulation
data relevant to single BHs with high accretion rates indicates
that $\sim 10\%$ of the heating takes place in regions optically thin
to Thomson scattering, and is therefore radiated as Compton-scattered
hard X-rays \citep{NK09,SchnittmanKN:2013}.  In our high accretion rate
case, the hard X-ray luminosity is a similar fraction of the mini-disk
luminosity.  As we have already remarked elsewhere, more realistic
emissivity models may lead to an augmentation of the X-ray luminosity.
In addition, because this simulation had a sizable cut-out at the
system center-of-mass, we were unable to see the ``sloshing" motion
described in \citet{Bowen17}; the shocks in this part of the flow
may also produce X-rays.

\subsection{Radiative Efficiency}
\label{sec:rad_eff}

Integrating over the face-on spectrum, we find that the emitted power
in the high accretion case
is $\simeq 0.1 L_{\rm Edd}$ rather than the $0.5 L_{\rm Edd}$ that
might be expected from $\dot m = 0.5$.  Several factors contribute to
this discrepancy (which applies equally well to the low accretion
rate case).  Roughly half can be attributed to our definition of
$\dot m$, which assumes 10\% radiative efficiency, whereas these are Schwarzschild BHs,
for which the canonical radiative efficiency, the binding energy at
the ISCO, is only 5.7\% (the actual radiative efficiency may reach $\gtrsim 6\%$
according to \citet{Noble11}).  A smaller decrease in the luminosity
observed in the face-on view is caused by relativistic effects: beaming
into the equatorial plane and gravitational redshift (\citet{Noble11}
found the polar suppression in a Schwarzschild spacetime to be $\simeq 10\%$).

In addition, as we have already emphasized, each mini-disk's outer edge is only twice as far from
its BH as its ISCO.  With specific angular momentum only slightly greater than that
at the ISCO, gas  can accrete with comparatively little in the way of dissipation.  When this
occurs, matter is accreted with greater orbital energy, and therefore radiates
almost a factor of two less efficiently \citep{IB01}.

Yet another relevant consideration is an artifact of the brief duration of our simulation.
To travel from the circumbinary disk to the binary, matter must go inward
once, be torqued to greater angular momentum, shock against the circumbinary
disk, lose angular momentum, and then fall toward the binary.  This
process takes roughly as long as an orbital period at the circumbinary
disk edge, $\simeq 3$ binary orbital periods.  Because our simulation
ran only three binary orbital periods, it did not run long enough for
the mass-supply rate to the binary to equilibrate with the mass accretion
rate in the inner portion of the circumbinary disk.

\subsection{Angular-Dependence and Time-Dependence} 

The accretion rate from a circumbinary disk to mini-disks surrounding
the individual members of a binary system is in general modulated on
frequencies comparable to the binary orbital frequency
\citep{MM08,Roedig:2011,Shi12,Noble12,Farris14}.  Whether this translates
into time-dependence of radiation depends on how swiftly the cross-gap
accretion rate translates into photon emission. When the inflow through
the mini-disks is governed by conventional disk mechanics (i.e., turbulent
MHD stresses), the inflow time is nearly always considerably longer than
the binary orbital period.   In the situation treated here, however, the
mini-disk inflow time is comparable to or shorter than the modulation period.
As mentioned in the previous subsection, this may in part be due to the
fact that the mini-disks' outer edges are not far from the ISCO, permitting
accretion to be driven by pressure gradients at least as much as by internal
stresses \citep{IB01}  Additionally, because
the mini-disks are in a binary, tidal forces induce spiral shocks that can
also transport angular momentum, particularly if the disk is comparatively hot
\citep{Lynden-Bell-Pringle74,Ju16}, and these are indeed present in our
simulation data \citep{Bowen18}.  As a result of the accelerated inflow
rate, the switching of the
accretion from one mini-disk to the other that accompanies the overall
modulation in accretion rate is reflected in a strong modulation of
the accretion rate in each mini-disk; if the inflow rate were slower,
the accretion rate in the individual mini-disks would track the longer-term
accretion rate, not the modulated version.

In this context, relativistic effects can create time-variability in a number
of ways.  As demonstrated in Section~\ref{sec:subeddington} in regard to the
low accretion rate case, Doppler beaming and gravitational lensing can work
together to induce periodic variation at twice the orbital frequency when
the system is viewed within $\approx 20^\circ$ of the orbital plane.
Although Figure~\ref{fig:lum_sphere} is posed in terms of azimuthal angle-dependence
in the corotating frame, this translates to time-dependence in an inertial frame.

Similar effects can also modulate the light observed from a more rapidly-accreting
system viewed not too far from the plane, but they work somewhat differently.
The key change is that these disks are much more optically thick.  This
means, for example, that they are more effective at blocking nearly edge-on
views.  In addition, and of greater interest, the time for photons to escape
from within a mini-disk can be comparable to or longer than the inflow time
within the mini-disk.  When the photon diffusion time is longer than the
inflow time, absorption can both smooth out the accretion rate modulation and also
suppress the emitted luminosity.  Quantitative evaluation of these effects
demands more powerful simulation tools.

\subsection{Caveats}
\label{sec:caveats}
\label{sec:caveats_fastlight}
\label{sec:caveats_acc_rate}
Our methods are based upon some approximations that are often, but not
always, valid.  One is the assumption of ``fast-light".  This
assumption essentially translates to ray-tracing within single
snapshots corresponding to a single value of coordinate time, even
though light requires a finite time to travel across the source region.
This approximation would not support predictions of time-variability
on scales comparable to or shorter than the light crossing-time.
However, because the photon diffusion time through optically-thick
disks is generically an order of magnitude or more longer than the
local dynamical time, and the dynamical time is $\sim (r/M)^{3/2}M$,
such rapid variability is unlikely to be important
for the thermal portion of the spectrum.  For the optically-thin case,
the relevant considerations are different.  Our model for the
Comptonized spectrum short-circuits any account of the time required
for the spectrum to reach a steady-state.  Because we set the
temperature in the optically thin corona component to a fixed value,
emulating the corona temperature seen in observed AGN, we neglect the time variability
associated with temperature fluctuations in the corona that may be
present in natural systems.  The variability in the corona from our
calculations is therefore due solely to the cooling time scale, set to
be the local orbital period---a time scale much larger than the local
light crossing time.  Thus, mainly because of the assumptions built
into our coronal emission model, our fast-light approximation is
expected to accurately capture the time variability of our system.

Since the fast-light approximation required us to use the
same metric time slice used by the simulation data snapshot, the
geodesic calculation also assumed the fast-light approximation.  The
spacetime dynamics are tied to the orbit of BHs, so if the ratio of
their orbital velocity to the speed of light is small then we would
expect that the fast-light approximation is justified.  At the
$a\simeq 20M$ separation considered here, this ratio is $\simeq 0.1$,
which is small enough for the approximation to be reasonable, but
only marginally so.  We will measure the error
of this assumption in future work in which we propagate the photons in
time with the spacetime and simulation data.

Another concern arises from the short duration of the underlying
simulation (3 binary orbits).  Over such a short time, the
system has not reached inflow equilibrium in any sense; put another
way, the accretion rate varies substantially as a function of radius.
This is likely true even if the accretion rates onto the mini-disks
are summed.
By contrast, because astrophysical BBHs have evolved for a number of orbits
many orders of magnitude larger than the 3 orbits of our GRMHD
simulation, in most instances they can be expected to have reached a very
close approximation to inflow equilibrium (in terms of the total
accretion rate onto the mini-disks).  Departures from a
radially-uniform accretion rate can lead to distortions in the
predicted spectrum by overemphasizing some temperature regions and
under-emphasizing others. In addition, as has been emphasized in
both this paper and \cite{Bowen18}, it is unclear to what degree
the mini-disk regions ever settle into a steady-state given the
possibility that they may follow a sequence of depletion/refilling
episodes if their inflow rates are comparable to or faster than
the binary orbital frequency.  If so, a time-dependent analysis
of their emission properties (as illustrated here) will be
necessary.

A third concern is that our simulation determines the local scale
height of the disk by a rather {\it ad hoc} mechanism.  In real disks,
it is the result of pressure support whose blend of radiation and gas
pressure, as well as its temperature, are determined by a balance
between turbulent dissipation and heat transport.  By contrast, in our
simulation pressure is assumed to be entirely gas pressure, and the
local temperature is held close to an imposed target temperature by
forcing the gas to lose heat at a rate comparable to the dynamical frequency.
Although these gross approximations may be very significant for
determination of the disk shape, they are likely less so for
time-averaged predictions of photon output.  One reason is that the
local rate of heat dissipation is dependent upon the mass accretion
rate and the gravitational potential, rather than the shape of the
disk. Another is that thermal equilibrium fixes the local spectrum
once the surface brightness (the dissipation rate per unit area) is
known.  Our very approximate description of the cooling rate has the
virtue of tying the local bolometric luminosity very closely to the
dissipation required by the local accretion rate.  It may, however, be
unreliable in regions (e.g., the streams or very optically thick
regions of the mini-disks) where the physical cooling
time may be long compared to the time required to change in radius.

A further concern, closely-related to the {\it ad hoc} cooling rate,
is that although we describe one case in which the accretion rate is
almost Eddington, we ignore both radiation forces and photon
trapping.  It is probably best to think of our $\mdothigh$ example
as an initial exploration of the properties of such disks.

It is also possible that by setting the criterion for emitting
a locally thermal spectrum to be the existence of a Thomson scattering
photosphere, we overestimated the thermal luminosity and underestimated
the unthermalized radiation.  Regions where the Thomson depth
is only $< O(10)$ may, in fact, produce spectra rather harder
than the Planckians we assigned them.

Our method (shooting rays from a distant camera to the source)
describes photon propagation well only when the entire path is
transparent.  It is essentially for this reason that we restricted our
high accretion rate predictions to face-on views.  For predictions
taking into account the shape of the photosphere, as well as Compton
scattering in coronal regions, a method that follows photons from
source to observer (e.g., {\it Pandurata}: \cite{2013ApJ...777...11S})
is preferable.

\section{Conclusions}
\label{sec:conclusion}
In this paper we presented a first step toward estimating the radiative
properties of SMBBHs in the stage immediately before merger.

When the accretion rate is great enough to make most of the accretion flow optically
thick, our model produces thermal radiation with a spectrum that differs only
modestly from ordinary single BH systems.  The contrasts may be
greater, however, for binaries with greater separations or containing
more rapidly-spinning BH or if regions of modest optical depth achieve
only partial thermodynamic equilibrium between gas and photons.

Outside thermalized regions, inverse Compton scattering between photons and high-energy
electrons produces hard X-ray emission.  
The hard X-ray flux may also be subject to modulation on
frequencies comparable to the binary orbital frequency, particularly
when the system is viewed from a position near the orbital plane and
the accretion rate is comparatively low.
Both Doppler beaming and gravitational lensing can modulate
the observed light flux seen by near-plane observers. Additional X-ray variability may arise from 
refilling/depletion episodes caused by periodic passage of the BHs
near the overdensity feature at the edge of the circumbinary disk.

Some of our predictions are robust; others are subject to cautions
we have enumerated in Section~\ref{sec:caveats}.  However, the post-processing tool we have created
has considerable flexibility and potential power, one that can be re-used---employing
more realistic assumptions---on data from future simulations.


\acknowledgments
S. D. was supported through the {\it Center for Computational Relativity and Gravitation} and  
{\it Frontier in Gravitational Astrophysics} program 
through RIT's office of research. D.~B., M.~C., V.~M. received support from NSF grants AST-1028087, AST-1516150,
PHY-1305730, PHY-1707946, OAC-1550436 and OAC-1516125.  S.~C.~N. was
supported by AST-1028087, AST-1515982 and OAC-1515969, and by an appointment to the
NASA Postdoctoral Program at the Goddard Space Flight Center
administered by USRA through a contract with NASA.  J.~H.~K. was
partially supported by NSF grants AST-1516299, PHYS-1707826 and
OAC-1516247 and the Simons Foundation (grant 559794, JHU).
V.M. also acknowledges partial support from AYA2015-66899-C2-1-P. 

Analysis and ray-tracing was performed on the Blue Waters system at the
University of Illinois at Urbana-Champaign and its National Center for
Supercomputing Applications, and the NewHorizons and BlueSky Clusters
at Rochester Institute of Technology.  The Blue Waters
sustained-petascale computing project is supported by the National
Science Foundation (awards OAC-0725070 and OAC-1238993) and the state
of Illinois. This work is also part of the ``Predicting the Transient
Signals from Galactic Centers: Circumbinary Disks and Tidal
Disruptions around Black Holes'' PRAC allocation support by the
National Science Foundation (award number OAC-151596).  The
NewHorizons and BlueSky Clusters were supported by NSF grant
No. PHY-0722703, DMS-0820923, AST-1028087, and PHY-1229173.  This work
was performed in part at Aspen Center for Physics, which is supported
by National Science Foundation grant PHY-1607611.

%
\pagebreak
\newpage

\appendix

\section{The Lorentz invariant radiative transfer equation}
\label{app:RTE}

Let $I$, $\alpha$, and $j$ be the Lorentz invariant 
intensity, absorption coefficient, and emissivity, respectively, and
let $I_\nu$, $\alpha_\nu$, and $j_\nu$ be the values measured by a local
observer that measures the photon to have frequency $\nu$. 
The relationship between these two sets of quantities is 
\begin{eqnarray}
I = \frac{I_\nu}{\nu^3} \quad , \quad 
j = \frac{j_\nu}{\nu^2} \quad ,  \quad 
\alpha = \alpha_\nu \nu \quad . \label{inv-variables} 
\end{eqnarray}
The radiative transfer equation in this observer's frame is 
\begin{equation}
\deriv{I_\nu}{s} \ = \ j_\nu - \alpha_\nu I_\nu \quad , \label{rad-trans}
\end{equation}
where $ds$ is the incremental distance along the geodesic as seen by the local observer.
Inserting Equations~(\ref{inv-variables}) into (\ref{rad-trans})
yields
\begin{equation}
\deriv{I}{s} \ = \ \frac{1}{\nu} \left( j - \alpha I \right) \quad . \label{rad-trans-2}
\end{equation}

We wish to use the affine geodesic parameter to evolve this
differential equation. If we denote the local observer's 4-velocity by
$v^\mu$ and the photon's wavevector by $k^\mu$, we have the following
relations:
\begin{eqnarray}
\nu = - \frac{v^\mu k_\mu}{2\pi} \quad , \label{nu} \\
ds = c \, dt_{obs} = - \frac{v^\mu dx_\mu}{c} \quad . \label{ds} \\
\end{eqnarray}
Using the freedom we have to choose $\lambda$ up to an affine
transformation, we define the normalized wavevector $N_\mu
= \pderiv{x^\mu}{\lambda} \equiv \frac{c}{2\pi} k^\mu$. Equations
(\ref{nu}) and (\ref{ds}) yield $ds = \nu d\lambda$, and we may
rewrite (\ref{rad-trans-2}) in the simple form:
\begin{equation}
\deriv{I}{\lambda} = j - \alpha I \quad . \label{rad-trans-inv}
\end{equation}

\section{Conversion factors}
\label{app:conversion}

In order to convert variables used in \harm and \bothros to cgs units, we need to match numerical scales to cgs scales.   In the following, we subscript with an ``n'' the variables in numerical units, with a ``c'' those in cgs units and with ``c/n'' the conversion factors between the two system of units, $X_{c/n} \equiv X_c / X_n$. Let us now discuss how $\nu_c$ is calculated from $\nu_n$.  We have:
\begin{equation}
\nu_{c/n} = - \frac{\left(u_\mu N^\mu\right)_{c/n}}{ c } .
\label{freq-conversion}
\end{equation}

Since the conversion factors should all be constants along the geodesics, we can evaluate them at the camera.  We denote by $C^\mu$ the $4$-velocity of the camera, which we set to $(1,0,0,0)$, and the subscript $\infty$ denotes a quantity evaluated at the camera. If the camera observes the photon at a given frequency $\nu_{c \, \infty}$, we then get:
\begin{equation}
\left(u_\mu N^\mu\right)_{c/n} 
= \frac{\left(C_\mu N^\mu\right)_{c \, \infty}}{
        \left(C_\mu N^\mu\right)_{n \, \infty}}
= - \frac{ c \, \nu_{c \, \infty}  }{ \left(C_\mu N^\mu\right)_{n \, \infty} } 
\quad . 
 \label{v-N-conversion}
\end{equation}
As we mentioned in Section~\ref{sec:geodesic}, $N^\mu_\infty$ is set
such that the geodesics are null ($N^\mu N_\mu = 0$), point
to the camera from the field, and move forward in time.
Putting this into (\ref{freq-conversion}), we obtain:
\begin{equation}
\nu_{c/n} = - \frac{\left(u_\mu N^\mu\right)_{c/n}}{ c }
= - \frac{ \nu_{c \, \infty}  }{ \left(C_\mu N^\mu\right)_{n \, \infty} } \quad .
\label{nu-conversion}
\end{equation}

The density scale of the gas is set by
choosing an average accretion rate into the system normalized by the numerical
one measured from the simulation data.  Because our
circumbinary disk has reached inflow equilibrium only out to $r \sim
4a$ we choose $2a < r < 4a $ as the region within which to measure the
simulation's accretion rate in code units, i.e.  $\dot{M}_n$.
Measuring this from the \cite{Noble12} simulation, we find
$\dot{M}_n \approx 0.03$ at $t=50,000M$, the time from which we start
the \cite{Bowen18} simulation.

The other quantities are more straightforward to transform, and the scales used are shown in Table~\ref{table:scales}.

\begin{table}[htb]
\begin{center}
\begin{tabular}{|r|c|c|c|}
\hline
Quantity ($Q$) &Numerical ($Q_n$) &CGS ($Q_c$) &Conversion ($Q_{c/n}$) \\
\hline \hline
Length ($L$) &$M_n \ [1]$   &$G M_c / c^2 $ &$L_c/L_n$\\[0.1cm]
Time   ($T$) &$M_n \ [1]$   &$c L_c $ &$c L_{c/n}$ \\[0.1cm]
Mass   ($M$) &$M_n \ [1]$ &$M_c \ [10^6 M_\odot]$ &$M_c/M_n$ \\[0.1cm]
Accret. Rate ($\dot{M}$) &$\dot{M}_n \ [0.03]$ 
                                &$\dot{M}_c \ [\mdotlowvalue - \mdothighvalue~\dot{M}_{\mathrm{Edd}}]$ 
                                &$\dot{M}_c/\dot{M}_n$ \\[0.1cm]
Mass Density ($\rho$) &$\dot{M}_n/\left(4\pi L_n^2\right) $    &$\dot{M}_c / \left(4\pi c L_c^2 \right)$
                      &$ \dot{M}_{c/n} / \left( c L_{c/n}^2 \right)$\\[0.1cm]
Frequency ($\nu$) &$-\left(C_\mu N^\mu\right)_{n \, \infty} [1]$ 
                  &$\nu_{c \, \infty} \ [10^{14} - 10^{21} \mathrm{Hz}]$ 
                  &$ - \nu_{c \, \infty} / \left(C_\mu N^\mu\right)_{n \, \infty}$\\[0.1cm]
Affine Parameter ($\lambda$) &$L_n / \nu_n$ & $L_c / \nu_c$ &$L_{c/n} / \nu_{c/n} $\\[0.1cm]
\hline
\end{tabular}
\end{center}
\caption{Scales used in \bothros to determine the conversion factors in translating 
numerical units to cgs units. Please refer to
Equation~(\ref{nu-conversion}) regarding $\nu_n$. $\dot{M}_n$ is
measured at the inner edge of the circumbinary.  We see that all
conversion factors can be derived from the free parameters $M_c$, $\dot{M}_c$ and
$\nu_{c \, \infty}$.  Actual scaling parameters used are specified in square brackets ``[]''.
\label{table:scales}  }
\end{table}

\pagebreak
\newpage
\bibliography{References}

\end{document}